\documentclass{article}

\usepackage{arxiv}
\usepackage{xcolor}
\usepackage[utf8]{inputenc} 
\usepackage[T1]{fontenc}    
\usepackage{hyperref}       
\hypersetup{
	colorlinks=true,
	linktoc=all,
	citecolor=darkgray,
	linkcolor=darkgray,
	urlcolor=darkgray
}

\usepackage{url}            
\usepackage{booktabs}       
\usepackage{amsfonts}       
\usepackage{nicefrac}       
\usepackage{microtype}      
\usepackage{lipsum}		
\usepackage{graphicx}
\usepackage[square,comma,numbers]{natbib}
\usepackage{doi}
\usepackage{amsmath}
\usepackage{amssymb}    
\usepackage{bm}
\usepackage{upgreek}
\usepackage{relsize}
\usepackage[font=small,labelfont=bf,figurename=Fig.]{caption}
\usepackage{subcaption}
\usepackage{booktabs}
\usepackage{mathtools}

\usepackage[nameinlink,capitalize]{cleveref}
\usepackage[linesnumbered,ruled,vlined]{algorithm2e}

\DeclareMathAlphabet\mathbfcal{OMS}{cmsy}{b}{n}
\DeclareMathOperator{\vect}{vec}
\DeclareMathOperator{\diag}{diag}

\newcommand{\?}{\:\!}

\title{Sequential Bayesian Inference for Uncertain Nonlinear Dynamic Systems: A Tutorial}

\author{ 
Konstantinos E. Tatsis \\
	Institute of Structural Engineering\\
	ETH Zurich\\
	Zurich, Switzerland\\
	\texttt{tatsis@ibk.baug.ethz.ch} \\
	\And
	Vasilis K. Dertimanis \\
	Institute of Structural Engineering\\
	ETH Zurich\\
	Zurich, Switzerland\\
	\texttt{v.derti@ibk.baug.ethz.ch} \\
	\And
	Eleni N. Chatzi \\
	Institute of Structural Engineering\\
	ETH Zurich\\
	Zurich, Switzerland\\
	\texttt{chatzi@ibk.baug.ethz.ch} \\
}

\renewcommand{\shorttitle}{}

\hypersetup{
pdftitle={Sequential Bayesian Inference for Uncertain Nonlinear Dynamic Systems: A Tutorial},
pdfauthor={Konstantinos E. Tatsis},
pdfkeywords={Bayesian filtering, nonlinear non-Gaussian estimation, sequential Monte Carlo, importance sampling, Rao-Blackwellised Particle filtering},
}

\definecolor{edits}{rgb}{0, 0, 0}

\begin{document}
\maketitle

\begin{abstract}
	In this article, an overview of Bayesian methods for sequential simulation from posterior distributions of nonlinear and non-Gaussian dynamic systems is presented. The focus is mainly laid on sequential Monte Carlo methods, which are based on particle representations of probability densities and can be seamlessly generalized to any state-space representation. Within this context, a unified framework of the various Particle Filter (PF) alternatives is  presented for the solution of state, state-parameter and input-state-parameter estimation problems on the basis of sparse measurements. The algorithmic steps of each filter are thoroughly presented and a simple illustrative example is utilized for the inference of i) unobserved states, ii) unknown system parameters and iii) unmeasured driving inputs.
\end{abstract}


\section{Introduction}

The	presence of diverse energy dissipation mechanisms in structural systems is directly connected to nonlinear phenomena, which are typically triggered upon exceedance of the material capacity or small displacement regime. Therefore, nonlinear behavior is oftentimes observed in dynamic systems, which are subjected to significantly high levels of excitation, such as earthquakes and wind fields. As a result, the nonlinearities are associated, for the vast majority of structural and mechanical systems, with the restoring force components, in which case the governing equations of motion are classified as nonlinear.

\color{edits}
In general, however, the treatment of dynamics forms a challenging task, even when linear, particularly 
\color{black}
under the constraint of real-time performance, which is typically the case in Structural Health Monitoring (SHM) and control applications. Within this context, numerous time-domain methodologies have been proposed for system identification including the more simplified constructs that find application to linear systems, such as autoregressive moving average models (ARMA), the least squares estimation (LSE) method \citep{Smyth2002,Nagarajaiah2004}, the Eigensystem Realization Algorithm (ERA) \citep{Juang1985}, as well as a large class of subspace identification techniques \citep{Overschee1994,Peeters1999,Katayama2005}. 
\color{edits}
A particular class of identification schemes, relies on adoption of an observer setup, often in the form of
\color{black}
Bayesian filtering methods, such as the Observer/Kalman filter Identification (OKID) \citep{Juang1993,Fraraccio2008}, which aim at identifying the system dynamics by recursively, or even in batch mode \citep{Ebrahimian2017}, assimilating data into the underlying model structure of the system at hand.
\color{edits}
These filtering formulations are particularly fitting within the context of virtual sensing, i.e., the task of inferring response quantities at unmeasured locations \citep{Tchemodanova2021,Vettori2021}, or even unknown system properties. Virtual sensing is essential for tasks such as digital twinning, diagnostics of condition in critical yet unreachable locations and control.
\color{black}

\color{edits}
The estimation task increases in complexity for the case of nonlinear systems. A nonlinear formulation can arise even for the case of linear dynamics, when a joint state and parameter estimation task is pursued, as will be elaborated in this work.
\color{black}
Perhaps the simplest approach to tackle nonlinearities consists in successive linearizations of the system equations, leading to the so-called extended Kalman filter \citep{Ebrahimian2015,Wan2018}. A number of contributions based on these filters has been established for estimating not only the state and parameters (e.g. stiffness, damping or hysteretic properties) of a dynamic system, but also the unknown inputs \citep{Naets2015,Maes2019a,Rogers2020a}. Alternatively, particle-based methods \citep{Wu2007,Chatzi2010}, with the Unscented Kalman filter (UKF) constituting an efficient and widely used option \citep{Astroza2015,Erazo2018} among them, can operate directly on the nonlinear equations of motion and propose a sample- or particle-based representation of the 
\color{edits}
sequence of 
\color{black}
probability densities \citep{Chen2005,Ching2006}. These approaches are also known as sequential Monte Carlo samplers \citep{Doucet2000a,Moral2006}, such as the standard Particle Filter (PF), which have been successfully applied to both structural and complex mechanical systems \citep{Erazo2017,Tatsis2018}. One of the major drawbacks related with the techniques is the so-called sample impoverishment, which has been addressed in the literature by different means, such as evolutionary algorithms \citep{Kwok2005,Park2007,Chatzi2013}, extensions to more well-conditioned formulations \citep{VanderMerwe2004} and marginalized samplers \citep{Goodall2005,Olivier2017a}.

The implementation of these filters is based on the approximation of the posterior state distribution by generating a number of samples, the so called particles, using Monte Carlo methods \citep{Arnaud2001,Sarkka2010}. Particle filters constitute essentially the extension of point-mass filters with the difference that the particles are concentrated in regions of high probability, instead of being uniformly distributed. These filters though, do not only suffer from the sample impoverishment problem, which pertains to the loss of particle diversity in time, but are also greedy in terms of particles. This implies that their performance might quickly become computationally inefficient due to the large pool of particles required for the distribution approximations. This issue can become even more pronounced in the case of joint state and parameter estimation problems \citep{Chatzi2013}, where the state vector is typically augmented in order to include the sought-after parameters.

\color{edits}
At this point, it is worth mentioning a further identification task that is common within the SHM context, namely estimation under unknown input.
\color{black}
Several Bayesian approaches have been proposed for jointly estimating the unmeasured input and the partially observed state of both linear \citep{Tatsis2016,Nayek2019} and nonlinear systems \citep{Lei2019,Rogers2020}. In the case of linear systems, an unbiased minimum-variance recursive filter without direct transmission has been initially proposed in \cite{Kitanidis1987a} and later reformulated in \cite{Hsieh2009}. To address the lack of optimality of these estimators in terms of the mean squared error, a new more effective filter has been developed in \cite{Gillijns2007,Gillijns2007a} for joint input and state estimation. The numerical instabilities occurring therein, which are mainly observed when the number of measurements is smaller than the order of the system, have been also recently addressed in \cite{Lourens2012c}. The state augmentation technique, which is mainly adopted for the joint identification of state and parameters, has been also employed in structural systems for state and input estimation \citep{Lourens2012b,Naets2015b}. In a similar context, a dual Kalman filter (DKF) was derived in \citep{EftekharAzam2015,Azam2015}, which seems to resolve the numerical issues related to the AKF. The required invertibility and observability conditions for all these joint estimators have been explored and reported in \citep{Maes2015a}. The same authors have also recently proposed the use of a smoothing algorithm \citep{Maes2018}, which aims at reducing the input-state estimation uncertainty by introducing a small time delay in the output.

\color{edits}
A known issue
\color{black}
associated with use of the above-mentioned Bayesian-type filters lies in the tuning of the process and measurement noise covariance matrices \citep{Smith2009}, 
\color{edits}
which can severely affect performance when parameter estimation is pursued. This primarily affects estimation regimes, where the parameters are simultaneously estimated with the state, within the filtering setting, as opposed to schemes that employ a separate step for parameter estimation (e.g. Expectation-Maximization \citep{Chitralekha2009}). In joint state-parameter estimation,
\color{black}
the quality of inference is strongly related to the initial parameter values and cross-covariance terms, which are responsible for translating the information of observed states into corrections of the hidden parameters. It has been demonstrated that a constant value for the coupling terms of the process noise covariance might prove insufficient for these type of problems \citep{Smith2009}, since a flow-dependent tuning \citep{Carrassi2011} is required in order to avoid instabilities \citep{Yang2009,Koyama2010}. Therefore, due to the fact that the actual error statistics of these predictors are unknown, the robust and reliable estimation might prove to a significant challenge. For this purpose, a number of contributions has been published on the adaptation of the disturbance terms \citep{Mehra1970,Myers1976}, which can be classified into Baysian, maximum-likelihood and covariance matching approaches \citep{Tatsis2020,Odelson2006}.

On the other hand, the main advantage of all aforementioned Bayesian filters lies in the recursive implementation, which allows for online and real-time performance. To accomplish the latter, the system dynamics ought to be typically represented in a reduced-order space, in order to deliver computationally efficient evaluations. Such models have been successfully used with Kalman-type filters \citep{Lourens2012c,Maes2015a} however, the need for computationally efficient models becomes even more critical and also more challenging in the case of state-parameter and input-state parameter problems, for which reduced-order representations and system parametrizations have been proposed only for simple educational examples, such as beam models \citep{Naets2015}, spring-mass-dampers \citep{Dertimanis2019} and shear frames \citep{Maes2019a}. In this sense, the real-time implementation of such schemes with applications to more realistic and challenging systems calls for more sophisticated reduced-order representations \citep{Tatsis2018,Tatsis2021,Tatsis2022}, which can account for more complex system dynamics and parameter dependencies.

The structure of this paper, whose aim is to provide an overview of particle-based Bayesian filters for nonlinear non-Gaussian systems, is organized as follows: the problem of Bayesian filtering for stochastic state estimation in nonlinear system is presented in \cref{sec:Bayes} in terms of the Chapman-Kolmogorov and Bayes equations. In \cref{sec:state-parameter}, the problem is further generalized to joint state and parameter estimation problems and the sampling-based algorithms for optimal Bayesian filtering, namely the Unscented Kalman Filter (UKF), the Particle Filter (PF), the Particle Filter with Mutation (MPF), the Sigma-Point Particle Filter (SPPF), the Gaussian-Mixture Sigma-Point Particle Filter and the Rao-Blackwelised Particle Filter (RBPF) are presented. The problem is further extended to input-state and parameter estimation problems in \cref{sec:state-input-parameter}, where the recently propose Dual Kalman Filter - Unscented Kalman Filter is summarized. Lastly, the performance of the algorithms is presented in \cref{sec:applications} by means of an academic example, which we deem necessary for comprehensive illustration of state, state-parameter and input-state-parameter estimation cases.


\section{Problem Formulation}
\label{sec:Bayes}

Let us represent generic dynamic systems via use of the following general and nonlinear process equation in the continuous-time domain

\begin{equation}\label{eq:state}
	\dot{\mathbf{x}}(t) = \mathbf{F}\big(\mathbf{x}(t), \mathbf{p}(t)\big)+\mathbf{v}(t)
\end{equation}

where $\mathbf{x}(t)\in\mathbb{R}^n$ is the state vector, $\mathbf{p}(t)\in\mathbb{R}^{n_{\mathrm{p}}}$ is the input force vector, $\mathbf{F}:\mathbb{R}^{n+n_{\mathrm{p}}}\to\mathbb{R}^n$ denotes the vector-valued function that describes the system dynamics, 
\color{edits}
and which in the structural context may encompass geometric (due to large deformations or instabilities) or material (e.g. due to plasticity) nonlinearities,
\color{black}
and $\mathbf{v}(t)\in\mathbb{R}^n$ is the zero mean process noise vector with covariance matrix $\mathbf{Q}(t)\in\mathbb{R}^{n\times n}$. The corresponding observation equation, which is herein formed directly in the discrete-time domain reads

\begin{equation}\label{eq:obs}
	\mathbf{y}_{k} = \mathbf{H}\left(\mathbf{x}_{k},\mathbf{p}_k\right)+\mathbf{w}_{k}
\end{equation}

where $\mathbf{y}_k\in\mathbb{R}^{n_{\mathrm{y}}}$ is the observation vector at $t=k\Delta t$, $\mathbf{H}:\mathbb{R}^n\to\mathbb{R}^{n_{\mathrm{y}}}$ denotes the observation function, $\mathbf{x}_{k}\in\mathbb{R}^n$ and $\mathbf{p}_k\in\mathbb{R}^{n_{\mathrm{p}}}$ are the discrete-time equivalents of the state $\mathbf{x}(t)$ and input $\mathbf{p}(t)$ respectively at $t=k\Delta t$,  and $\mathbf{w}_{k}\in\mathbb{R}^{n_{\mathrm{y}}}$ is the zero-mean observation noise vector, whose covariance matrix is denoted by $\mathbf{R}_{k}\in\mathbb{R}^{n_{\mathrm{y}}\times n_{\mathrm{y}}}$. \cref{eq:state} can further be expressed in the discrete-time domain, thus yielding the following nonlinear discrete-time state-space model

\begin{align}
	\mathbf{x}_{k+1} & = \mathbf{G}\left(\mathbf{x}_{k},\mathbf{p}_{k}\right)+\mathbf{v}_{k}
	\label{eq:state-2}\\[2mm]
	\mathbf{y}_{k} & = \mathbf{H}(\mathbf{x}_{k},\mathbf{p}_k)+\mathbf{w}_{k}
	\label{eq:obs-2}
\end{align}

where $\mathbf{v}(t)$ and $\mathbf{Q}(t)$ are also mapped to the discrete-time counterparts $\mathbf{v}_k$ and $\mathbf{Q}_k$, respectively. 
\color{edits}
In the SHM context, the observations pertain to typically measurable quantities, including accelerations, velocities, tilts or strains. The covariance of the process, $\mathbf{Q}_k$, and measurement noise, $\mathbf{R}_k$, are tunable filter parameters, which reflect the confidence that may be attributed to the fidelity of the model and the precision of acquired measurements, respectively. The relative selection of these parameters affects both the convergence and persistent error bounds of the estimation process, as elaborated in \cite{Rhudy2013} for the case of the linear KF.
\color{black}
Similarly, the system dynamics function $\mathbf{G}:\mathbb{R}^{n+n_{\mathrm{p}}}\to\mathbb{R}^n$ is obtained upon integration of $\mathbf{F}$, as follows

\begin{equation}\label{eq:integ}
	\mathbf{G}(\mathbf{x}_{k}, \mathbf{p}_{k}) = \mathbf{x}_{k}+\int_{k\Delta t}^{(k+1)\Delta t}\mathbf{F}\big(\mathbf{x}(t),\mathbf{p}(t)\big)\,dt
\end{equation}

From a Bayesian point of view, the problem of determining posterior estimates of the state $\mathbf{x}_{k|k}$, i.e., estimates at time $k$ given the sequence of all available measurements $\mathbf{Y}_{k}=\left[\mathbf{y}_0\ \,\mathbf{y}_1\ \,\ldots\ \,\mathbf{y}_k \right]$ up to time $k$, may be tailored to a recursive scheme, whereby the prior $p(\mathbf{x}_{k}|\mathbf{Y}_{k-1})$ and posterior $p(\mathbf{x}_{k}|\mathbf{Y}_{k})$ distributions are obtained sequentially at each time instant $k$. Assuming the prior distribution $p(\mathbf{x}_0)$ is known and the required PDF $p(\mathbf{x}_{k-1}|\mathbf{Y}_{k-1})$ at time $k-1$ is also available, the prior probability $p(\mathbf{x}_k|\mathbf{Y}_{k-1})$ can be obtained through the prediction step, which is essentially derived from the Chapman-Kolmogorov equation

\begin{equation}\label{eq:prediction}
	\ p(\mathbf{x}_k|\,\mathbf{Y}_{k-1}) = \int{p(\mathbf{x}_k|\,\mathbf{x}_{k-1})p\left(\mathbf{x}_{k-1}|\,\mathbf{Y}_{k-1}\right)\,d\mathbf{x}_{k-1}}
\end{equation}

The probabilistic model of the state evolution $p\left(\mathbf{x}_k|\mathbf{x}_{k-1}\right)$, which is also referred to as transitional density, is defined by the process equation \cref{eq:state-2}, namely the system dynamics function $\mathbf{G}(\mathbf{x}_k)$ and the distribution $p(v_{k})$ of the process noise. Subsequently, the prior may be conditioned on the measurement $\mathbf{y}_k$ at time $k$, using Bayes Theorem as follows

\begin{equation}\label{eq:update}
	p\left(\mathbf{x}_k|\mathbf{y}_{k}\right) = p\left(\mathbf{x}_k|\mathbf{y}_k,\mathbf{Y}_{k-1}\right) = \frac{p(\mathbf{y}_{k}|\mathbf{x}_k)\,p(\mathbf{x}_k|\mathbf{Y}_{k-1})}{p(\mathbf{y}_k|\mathbf{Y}_{k-1})} 
\end{equation}

where the denominator $p(\mathbf{y}_k|\mathbf{Y}_{k-1})$, which acts as a normalizing factor, depends on the likelihood function $p(\mathbf{y}_k|\mathbf{x}_k)$, which is linked to the measurement process given by \cref{eq:obs-2}. The latter is in turn specified by the observation function $\mathbf{H}(\mathbf{x}_k)$ and the distribution of the measurement noise $\mathbf{w}_{k}$.

\color{edits}
The relevance of Bayesian inference in structural dynamics is primarily applicable to model-based Structural Health Monitoring (SHM) applications. Within such a context, the aim is to infer the dynamic state of structural systems by fusing a numerical system representation, which specifies the state process as per \cref{eq:state-2}, with sparse sensory information, whose type and location on the physical system define the measurement process expressed by \cref{eq:obs-2}. This inference step is oftentimes required to perform in real time, so as to allow for online assessment of the system dynamics, thus calling for continuous and sequential conditioning of the model prediction on the vibration-based measurement data.

\color{black}	

The recursive prediction and correction formulas postulated by \cref{eq:prediction,eq:update} form the basis of the optimal Bayesian solution. Once the posterior PDF is computed, the optimal state estimate can be obtained by means of different statistical metrics. In engineering applications, this is typically the conditional mean of $\mathbf{g}:\mathbb{R}^n\to\mathbb{R}^m$, which can be any arbitrary function with respect to $\mathbf{x}_k$, and represents the minimum square error (MMSE) estimate

\begin{equation}\label{eq:condmean}
	\mathbb{E}\left[\mathbf{g}\left(\mathbf{x}_{k}\right)\big|\mathbf{Y}_{k}\right] = \int{\mathbf{g}\left(\mathbf{x}_{k}\right)\, p\left(\mathbf{x}_k\big|\mathbf{Y}_{k}\right)\,d\mathbf{x}_k}
\end{equation}

\color{edits}
The function $\mathbf{g}$ is a user-defined parameter, linked to the requirements of the problem at hand, where often the quantities of interest that are to be inferred are specified as functions of the state vector $\mathbf{x}_k$. In control applications, the target engineering quantity is typically the state itself, either in terms of displacements or in terms of velocities. This implies that $\mathbf{g}(\mathbf{x}_k) = \mathbf{x}_k$ and as such the expectation of \cref{eq:condmean} is essentially the first moment of the posterior distribution $p\left(\mathbf{x}_k\big|\mathbf{Y}_{k}\right)$. The same applies to those applications which aim at inferring the entire response field using a limited number of observations. In SHM applications, where the aim might be for instance the inference of fatigue damage accumulation at unmeasured system locations \cite{Papadimitriou2011}, the quantity of interest is the stress field and as such, the function $\mathbf{g}$ is responsible for transforming the state variable $\mathbf{x}_k$ to stress values at fatigue-critical locations. It should be noted that the quantities of interest may be alternatively inferred by means of the maximum \textit{a posteriori} (MAP) estimate, which corresponds to the value of $\mathbf{g}\left(\mathbf{x}_k\right)$ that maximizes $p\left(\mathbf{x}_k|\mathbf{Y}_k\right)$ or even higher order statistics.

\color{black}

The prediction and filtering steps described by \cref{eq:prediction,eq:update} are amenable to an analytical solution only in the particular case of linear system equations and additive Gaussian noise. In the case of nonlinear system dynamics, which fall within the scope of this paper, the solution is approximated using the sampling algorithms described in the following sections. It should be mentioned for the sake of generality that the Extended Kalman Filter is another alternative for dealing with nonlinear Gaussian problems, requiring though successive linearizations at each time step, around the posterior estimates. This might prove to be computationally inefficient for real-time or near real-time performance and as such, the focus of this paper is laid on the sample-based approximations of the filtering equations.

\section{State and parameter estimation}
\label{sec:state-parameter}

\color{edits}
As a first instance of nonlinear problems, we here present the class of state-parameter estimation problems. This is a typical pursuit in the context of SHM, when the monitored system features a model of known structure, ,albeit of uncertain (or unknown parameters (e.g. stiffness, damping, or hysteretic parameters). The original
\color{black} Bayesian problem, which is formulated in \cref{sec:Bayes}, which essentially assumes that the state-space model is \textit{a priori} fully specified. However, as aforementioned, certain model parameters, which will be henceforth denoted by $\bm{\theta}\in\mathbb{R}^{n_{\theta}}$, might be unknown as well. As such, their estimation within the sequential Bayesian context can be performed by treating them as random variables, for which a prior distribution $p\left(\bm{\theta} \right)$ need be specified. The initial problem is therefore modified as follows
\begin{align}
	\bm{\theta}_k & \sim \mathcal{N}\left(\bm{\theta}_k;\bm{\theta}_{k-1},\bm{\Sigma}_{k-1}^{\,\theta} \right)\\
	\mathbf{x}_{k+1} & = \mathbf{G}\left(\mathbf{x}_{k}, \mathbf{p}_k, \bm{\theta}_k \right)+\mathbf{v}_{k}\\[1mm]
	\mathbf{y}_k & = \mathbf{H}\left(\mathbf{x}_k,\mathbf{p}_k,\bm{\theta}_k \right)+\mathbf{w}_k
\end{align}

\color{edits}
where the transition model of the parameters is assumed to be a Gaussian random walk, derived from the density $\mathcal{N}\left(\mathbf{x}; \mathbf{m}, \mathbf{C}\right)$ with argument $\mathbf{x}\in\mathbb{R}^n$, mean $\mathbf{m}\in\mathbb{R}^n$ and covariance $\mathbf{C}\in\mathbb{R}^{n\times n}$.
\color{black}
A straightforward way of integrating the parameter estimation problem into the Bayesian framework is to use the so-called state augmentation approach, whereby the parameter vector is appended to the system state. This implies that the state vector $\mathbf{x}_{k}$ is redefined as $\tilde{\mathbf{x}}_k = \vect\left(\left[\mathbf{x}_k, \bm{\theta}_k\right]\right)$ and the state-space model is now written in the augmented form
\begin{align}
	\tilde{\mathbf{x}}_{k+1} & =\tilde{\mathbf{G}}\left(\tilde{\mathbf{x}}_{k},\mathbf{p}_k \right)+\tilde{\mathbf{v}}_k\label{eq:augmented-state}\\[1mm]
	\mathbf{y}_k & = \tilde{\mathbf{H}}\left(\tilde{\mathbf{x}}_k,\mathbf{p}_k \right)+\mathbf{w}_k\label{eq:augmented-output}
\end{align}
where the initial system and observation functions $\mathbf{G}\left(\bullet\right)$ and $\mathbf{H}\left(\bullet\right)$ are accordingly transformed to their augmented counterparts $\tilde{\mathbf{G}}\left(\bullet \right)$ and $\tilde{\mathbf{H}}\left(\bullet \right)$ and respectively. It should be noted that the sampling algorithms presented in the following sections may refer either to the initial problem described by \cref{eq:state-2,eq:obs-2}, or to the augmented one represented by \cref{eq:augmented-state,eq:augmented-output}. For the sake of simplicity though, all the steps will be formulated using the notation of the initial problem, as described by \cref{eq:state-2,eq:obs-2}.

\subsection{The Unscented Kalman Filter (UKF)}
\label{subsec:UKF}

The Unscented Kalman Filter (UKF) is a filtering algorithm, which draws its formulation from the Unscented Transform (UT) \cite{Wan2000,Julier2004} and boils down to the use of a structured set of points (sigma points) in order to approximate the mean and covariance of the target distribution.
\color{edits}
This is also the main difference between the UKF and the different variants of Particle Filters (PFs), which will be elaborated in the following sections; the former is based on deterministic sample points while the latter relies on Monte Carlo (MC) sampling for the evaluation of the expectation integrals.
\color{black}
In the context of the UKF, it is postulated that the target distribution is represented by a Gaussian density, whose hyperparameters are estimated using a number of deterministically selected sample points, also referred to as the sigma points. These are propagated through the nonlinear state-space equations, enabling thus reconstruction of the prior $p\left(\mathbf{x}_k|\mathbf{Y}_{k-1}\right)$ and posterior $p\left(\mathbf{x}_k|\mathbf{Y}_{k}\right)$ densities in terms of the sample-based mean and covariance. The UKF is related to the Bayesian formulation presented in the previous section through the following recursive formulas
\begin{align}
	p\left(\mathbf{x}_{k-1}|\mathbf{Y}_{k-1}\right) = \mathcal{N}\left(\mathbf{x}_{k-1};\hat{\mathbf{x}}_{k-1|k-1},\mathbf{P}_{k-1|k-1}\right)
	\label{eq:ukf-1}
	\\[1mm]
	p\left(\mathbf{x}_{k}|\mathbf{Y}_{k-1}\right) = \mathcal{N}\left(\mathbf{x}_{k};\hat{\mathbf{x}}_{k|k-1},\mathbf{P}_{k|k-1}\right)
	\label{eq:ukf-2}
	\\[1mm]
	p\left(\mathbf{x}_{k}|\mathbf{Y}_{k}\right) = \mathcal{N}\left(\mathbf{x}_{k};\hat{\mathbf{x}}_{k|k},\mathbf{P}_{k|k}\right)
	\label{eq:ukf-3}
\end{align}
\color{edits}
in which $\hat{\mathbf{x}}_{k-1|k-1}$ and $\mathbf{P}_{k-1|k-1}$ denote the state and error covariance estimates at step $k-1$ given measurements up to and including step $k-1$. 
It should be underlined that the approximation postulated by \cref{eq:ukf-1,eq:ukf-2,eq:ukf-3} is essentially identical to the one characterizing a general Gaussian filter. However, the inference integrals of a general Gaussian filter can be calculated in a closed form only for specific model structures. Instead, they are practically calculated using numerical approximations, like the one proposed by the UKF. Similar approximations can be obtained using the Gauss-Hermite Kalman Filter (GHKF) or the Cubature Kalman Filter (CKF), which are based on Gauss-Hermite quadrature and spherical cubature rules respectively for the numerical integration of the inference equations.
\color{black}

Assuming that the posterior state estimate $\hat{\mathbf{x}}_{k-1|k-1}$ at step $k-1$ and the corresponding covariance $\mathbf{P}_{k-1|k-1}$ are available, the \textit{a priori} state statistics at step $k$ can be calculated by propagating the sigma points $\mathbfcal{X}_{k-1|k-1}^j\in\mathbb{R}^n$ for $j=0,\ldots,2n$, through the state equation. These points are deterministically placed with respect to the posterior density hyperparameters as follows
%
%
\begin{equation}
	\mathbfcal{X}_{k-1|k-1}^j = 
	\begin{cases}
		\ \,\hat{\mathbf{x}}_{k-1|k-1}, & \quad j=0 \\[1mm]
		\ \,\hat{\mathbf{x}}_{k-1|k-1}+\sqrt{n+\lambda}\,\left[\!\sqrt{\mathbf{P}_{k-1|k-1}}\, \right]_j, & \quad j=1,\ldots,n \\[1.5mm]
		\ \,\hat{\mathbf{x}}_{k-1|k-1}-\sqrt{n+\lambda}\,\left[\!\sqrt{\mathbf{P}_{k-1|k-1}}\,\right]_{j-n}, & \quad j=n+1,\ldots,2n
	\end{cases}
	\label{eq:sigma-points}
\end{equation}
where $n$ is the state size and $\lambda=\alpha^2\left(n+\kappa \right)-n$ is a scaling parameter which depends on the constants $\alpha$ and $\kappa$. The former determines the spread of sigma points around the mean value and typically ranges between $10^{-4}$ and 1, while the latter acts as a secondary scaling parameter which is is usually equal to $3-n$. The reader is referred to \cite{Julier1995} for further details in this respect. The term $\left[\!\sqrt{\mathbf{P}_{k-1|k-1}} \right]_j$ denotes the $j\,$th column of the covariance matrix square root.

\begin{algorithm}[h]
	\small
	\SetAlgoLined
	\DontPrintSemicolon
	\SetKwInOut{Input}{Input}
	\SetKwInOut{Output}{Output}
	\SetKw{And}{and}
	
	\BlankLine
	Specify the parameters $\alpha$, $\beta$ and $\kappa$\; \vspace{1mm}
	
	Initialize the search distribution parameters $\hat{\mathbf{x}}_{\,0|0}$ and $\mathbf{P}_{\,0|0}$\;
	
	\BlankLine
	\For{$k = 1,\, 2,\, \ldots\, \,$}{
		Calculate sigma points $\mathbfcal{X}_{k-1|k-1}^{\,j}$ using \cref{eq:sigma-points}\;
		
		\BlankLine
		(a) Time update step\;
		\BlankLine
		
		Calculate prior sigma points $\mathbfcal{X}_{k|k-1}^{\,j}$ by propagating $\mathbfcal{X}_{k-1|k-1}^{\,j}$ through \cref{eq:state-2}\;\vspace{1mm}
		
		
		Obtain the prior state estimates $\hat{\mathbf{x}}_{k|k-1}$ and $\mathbf{P}_{k|k-1}$ using \cref{eq:UKF-prediction-1a,eq:UKF-prediction-1b}\;
		
		\BlankLine
		(b) Measurement update step\;
		\BlankLine
		
		Propagate the sigma points $\mathbfcal{X}_{k|k-1}^{j}$ through \cref{eq:obs-2} to obtain the predicted output $\hat{\mathbf{y}}_{k|k-1}$\;\vspace{1mm}
		
		Calculate the predicted output covariance $\mathbf{P}_{k}^{\mathrm{yy}}$ and Kalman gain using \cref{eq:obs3,eq:obs4,eq:obs5}\;\vspace{1.5mm}
		
		Obtain the posterior state estimates $\hat{\mathbf{x}}_{k|k}$ and $\mathbf{P}_{k|k}$ using \cref{eq:UKF-update-1a,eq:UKF-update-1b}\;
	}
	
	\caption{Unscented Kalman Filter (UKF)}
	\label{alg:hierarchical-filter}
\end{algorithm}

Each one of the posterior sigma points at step $k-1$ is propagated through the state equation, thus yielding a prior estimate of the sigma points $\mathbfcal{X}_{k|k-1}^j$ for $j=0, \ldots,2n$ at step $k$. These points represent the predicted density $p\left(\mathbf{x}_{k}|\mathbf{Y}_{k-1}\right)$, whose mean and covariance are approximated using a weighted sample mean and covariance
\begin{align}
	\hat{\mathbf{x}}_{k|k-1} & = \sum_{j=0}^{2n}{W_j^{\,\mathrm{m}}\mathbfcal{X}_{k|k-1}^j}\label{eq:UKF-prediction-1a}\\
	\mathbf{P}_{k|k-1} & = 
	\sum_{j=0}^{2n}W_j^{\,\mathrm{c}}
	\left[\mathbfcal{X}_{k|k-1}^j-\hat{\mathbf{x}}_{k|k-1}\right]
	\left[\mathbfcal{X}_{k|k-1}^j-\hat{\mathbf{x}}_{k|k-1}\right]^{\mathrm{T}}+\mathbf{Q}_{k-1}\label{eq:UKF-prediction-1b}
\end{align}
The mean and covariance weights, denoted by  $W_j^{\,\mathrm{m}}$ and $W_j^{\,\mathrm{c}}$ respectively, are constant and given by the following formulas
\begin{align}
	W_0^{\,\mathrm{m}} = \cfrac{\lambda}{\lambda+n},
	\hspace{5mm}
	W_0^{\,\mathrm{c}} = \cfrac{\lambda}{\lambda+n}+1-\alpha^2+\beta,
	\hspace{5mm}
	W_j^{\,\mathrm{m}} = W_j^{\,\mathrm{c}} = \cfrac{1}{2(n+\lambda)},\hspace{3mm}\mathrm{for}\hspace{3mm}
	j=1,\ldots,2n
\end{align}
where $\beta$ is a scalar parameter, which is related to the state distribution. For Gaussian distributions $\beta$ is set to 2, while the reader is referred to \cite{Julier1995} for further details. The predicted measurement is similarly obtained as the weighted average of the measurement equation evaluated for each one of the sigma points
\begin{equation}\label{eq:state6}
	\hat{\mathbf{y}}_{k|k-1} = \sum_{j=0}^{2n}{W_j^{\,\mathrm{m}}\,
	\mathbf{H}\left(\mathbfcal{X}_{k|k-1}^j,\mathbf{p}_{k}\right)}
\end{equation}
Thereafter, the measurement update equations are as follows:

\begin{align}
	\hat{\mathbf{x}}_{k|k} & = \hat{\mathbf{x}}_{k|k-1}+\mathbf{K}_{k}
	\left(\mathbf{y}_{k}-\hat{\mathbf{y}}_{k|k-1}\right)
	\label{eq:UKF-update-1a}\\[2mm]
	\mathbf{P}_{k|k} & = \mathbf{P}_{k|k-1}-\mathbf{K}_{k}^{\phantom{y}}\,\mathbf{P}_{k}^{\,\mathrm{yy}}\,\mathbf{K}_{k}^T
	\label{eq:UKF-update-1b}
\end{align}

where
\begin{align}
	& \mathbf{K}_{k} = \mathbf{P}_{k}^{\,\mathrm{xy}}
	\left(\mathbf{P}_{k}^{\,\mathrm{yy}}-\mathbf{R}_{k}\right)^{-1}
	\label{eq:obs3}\\[2mm]
	& \mathbf{P}_{k}^{\,\mathrm{yy}} = \sum_{j=0}^{2n}W_j^{\,\mathrm{c}}\left[\mathbf{H}\left(\mathbfcal{X}_{k|k-1}^j,\mathbf{p}_{k}\right)-\hat{\mathbf{y}}_{k|k-1}\right]\left[\mathbf{H}\left(\mathbfcal{X}_{k|k-1}^j,\mathbf{p}_{k}\right)-\hat{\mathbf{y}}_{k|k-1}\right]^{\mathrm{T}}+\mathbf{R}_{k}
	\label{eq:obs4}\\[1mm]
	& \mathbf{P}_{k}^{\,\mathrm{xy}} = \sum_{j=0}^{2n}W_j^{\,\mathrm{c}}\left[\mathbfcal{X}_{k|k-1}^j-\hat{\mathbf{x}}_{k|k-1}\right]\left[\mathbf{H}\left(\mathbfcal{X}_{k|k-1}^j,\mathbf{p}_{k}\right)-\hat{\mathbf{y}}_{k|k-1}\right]^{\mathrm{T}}
	\label{eq:obs5}
\end{align}
where $\mathbf{K}_{k}$ is the Kalman gain at step $k$. It should be noted that in the UKF implementation presented in this paper the effect of noise is simply additive, which is oftentimes encountered in engineering problems and provides a significant reduction of the computational complexity. In the case of non-additive noise, which implies that $\mathbf{G}(\bullet)$ and $\mathbf{H}(\bullet)$ are functions of the process and measurement noise terms respectively, the state vector need be augmented with the noise variables $\mathbf{v}_k$ and $\mathbf{w}_k$ and the sigma points need be calculated using the augmented state vector.
	


\subsection{The Particle Filter (PF)}
\label{subsec:PF}

In problems where Gaussian approximations are not suited for representing the target distribution, which is for instance the case when the latter is a multi-modal distribution, the filtering equations are solved with sequential Monte Carlo approximations, which are encapsulated in Particle Filters (PFs) \cite{Arulampalan2002}. Similar to the UKF, PFs approximate the target PDF $p\left(\mathbf{x}_k|\mathbf{Y}_k\right)$ using a set of sample points $\mathbf{x}_k^j$ for $j=1,\ldots,N$ with associated weights $w_k^j$. The major difference consists in the fact that a PF is based on samples drawn from a Monte Carlo scheme, while the UKF makes use of deterministically placed samples. However, due to practical difficulties in sampling directly from $p\left(\mathbf{x}_k|\mathbf{Y}_k\right)$, the expectation of \cref{eq:condmean} when using a PF is approximated by drawing samples from an importance distribution, which is herein denoted by $q(\mathbf{x}_k|\mathbf{Y}_k)$, per the so called importance sampling (IS) strategy. Within this context, the expectation over the probability density function $p\left(\mathbf{x}_k|\mathbf{Y}_k \right)$ can be decomposed as follows

\begin{equation}\label{eq:condmean2}
	\mathbb{E}\left[\mathbf{g}\left(\mathbf{x}_{k}\right)\big|\mathbf{Y}_{k}\right] = \int\mathbf{g}\left(\mathbf{x}_{k}\right)\, p\left(\mathbf{x}_k\big|\mathbf{Y}_{k}\right)\,d\mathbf{x}_k = 
	\int\left[\mathbf{g}\left(\mathbf{x}_k\right)\frac{p\left(\mathbf{x}_k|\mathbf{Y}_k\right)}{q\left(\mathbf{x}_k|\mathbf{Y}_k \right)} \right] q\left(\mathbf{x}_k|\mathbf{Y}_k \right) d\mathbf{x}_k
\end{equation}

on the condition that the importance distribution $q(\mathbf{x}_k|\mathbf{Y}_k)$ receives non-zero values at the non-zero points of $p\left(\mathbf{x}_k|\mathbf{Y}_k \right)$. As such, \cref{eq:condmean2} can be seen as the expectation of the terms contained within the brackets over the importance distribution, which can be sampled in order to obtain a Monte Carlo estimate of the integral.

Within this context, a Monte Carlo approximation of the expectation described by Eq. \eqref{eq:condmean2} can be obtained from the following weighted sum

\begin{equation}
	\mathbb{E}\left[\mathbf{g}\left(\mathbf{x}_{k}\right)\big|\mathbf{Y}_{k}\right] \approx \frac{1}{N}\sum_{j=1}^N \frac{p\left(\mathbf{x}_k^{\,j}\big|\mathbf{Y}_k\right)}{q\left(\mathbf{x}_k^{\,j}\big|\mathbf{Y}_k \right)}\,\mathbf{g}\left(\mathbf{x}_k^{\,j} \right)
	= \sum_{j=1}^N w_k^{\,j}\,\mathbf{g}\left(\mathbf{x}_k^{\,j} \right)
\end{equation}

where $w_k^{\,j}$ denotes the $j$-th sample weight. A similar expression can be derived for the approximation of the probability density $p\left(\mathbf{x}_k|\mathbf{Y}_k\right)$, in which case $\mathbf{g}\left(\mathbf{x}_k^{\,j}\right)$ would be substituted by $\delta\left(\mathbf{x}_k-\mathbf{x}_k^{\,j} \right)$. The importance weights can be generated by a recursive formula, which is derived \cite{Arnaud2001,Sarkka2010} by using the Markov properties of the importance distribution and the system model described by \cref{eq:state-2} and postulates the following proportionality 

\begin{equation}
	w_k^{\,j} \propto w_{k-1}^{\,j}\frac{p\left(\mathbf{y}_k\big|\mathbf{x}_k^{\,j} \right) p\left(\mathbf{x}_k^{\,j}\,\big|\,\mathbf{x}_{k-1}^{\,j} \right)}{q\left(\mathbf{x}_k^{\,j}\,\big|\,\mathbf{x}_{k-1}^{\,j},\mathbf{Y}_k \right)}
	\label{eq:weights-1}
\end{equation}

where \color{edits}
$p\left(\mathbf{x}_k^{\,j}\?\big|\?\mathbf{x}_{k-1}^{\,j} \right)$ 
\color{black}
is the transitional density, defined by \cref{eq:state-2}, and $p\left(\mathbf{y}_k\?\big|\?\mathbf{x}_k^{\,j} \right)$ is the likelihood function, which is fully specified by \cref{eq:obs-2}.

\begin{figure}[b]
	\centering
	\includegraphics[width=0.52\textwidth]{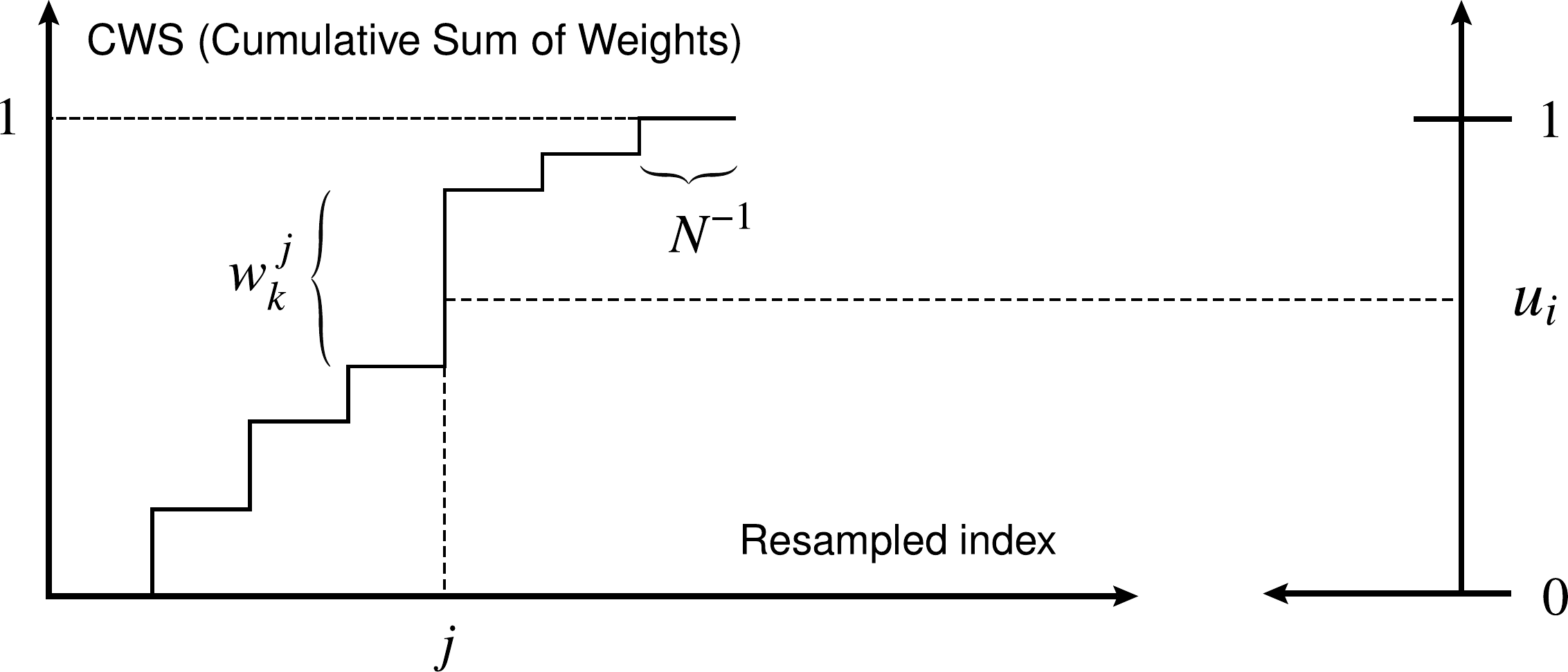}
	\caption{Schematic representation of the re-sampling step; the uniformly distributed random variable $u_i$ is mapped to index $j$ and the corresponding particle $\mathbf{x}_k^{\,j}$ is likely to be selected due to its considerable weight $w_k^{\,j}$}
	\label{fig:resample}
\end{figure}

As evidenced from the presented formulation, the performance of a PF is strongly dependent on the quality of the proposal distribution $q\left(\mathbf{x}_k|\mathbf{Y}_k \right)$, which should further allow to easily draw samples therefrom. It has been proved \cite{Doucet2000a} that the optimal importance distribution, in terms of variance, is $p\left(\mathbf{x}_k|\mathbf{x}_{k-1},\mathbf{y}_k \right)$. 
\color{edits}
However, a convenient selection that is oftentimes used is the transitional density $p\left(\mathbf{x}_k|\mathbf{x}_{k-1}^{\,j} \right)$, which results to the so called bootstrap particle filter. This is not necessarily a very informative function and not always an optimal one since the space of $\mathbf{x}_k$ is explored without using the information contained in the measurements. On the other hand, the transitional density can be more easily sampled and upon substitution in \cref{eq:weights-1} yields the following recursive formula for the calculation of weights \color{black}
\begin{equation}
	w_k^{\,j} = w_{k-1}^{\,j}\,p\left(\mathbf{y}_k\big|\mathbf{x}_k^{\,j} \right)
	\label{eq:weights-2}
\end{equation}
which implies that the selection of weights $w_k^{\,j}$ at time step $k$ depends on the likelihood $p\left(\mathbf{y}_k\big|\mathbf{x}_k^{\,j} \right)$ which is fully specified by the measurement process.
\color{edits}
Since the weights calculated by sampling a density function, be it the one of \cref{eq:weights-2} or in the more general case the one of \cref{eq:weights-1}, represent the likelihood of each particle, they should be normalized so that the sum of all likelihoods is equal to one.
\color{black}

\begin{figure}[t!]
	\begin{center}
		\includegraphics[width=0.8\textwidth]{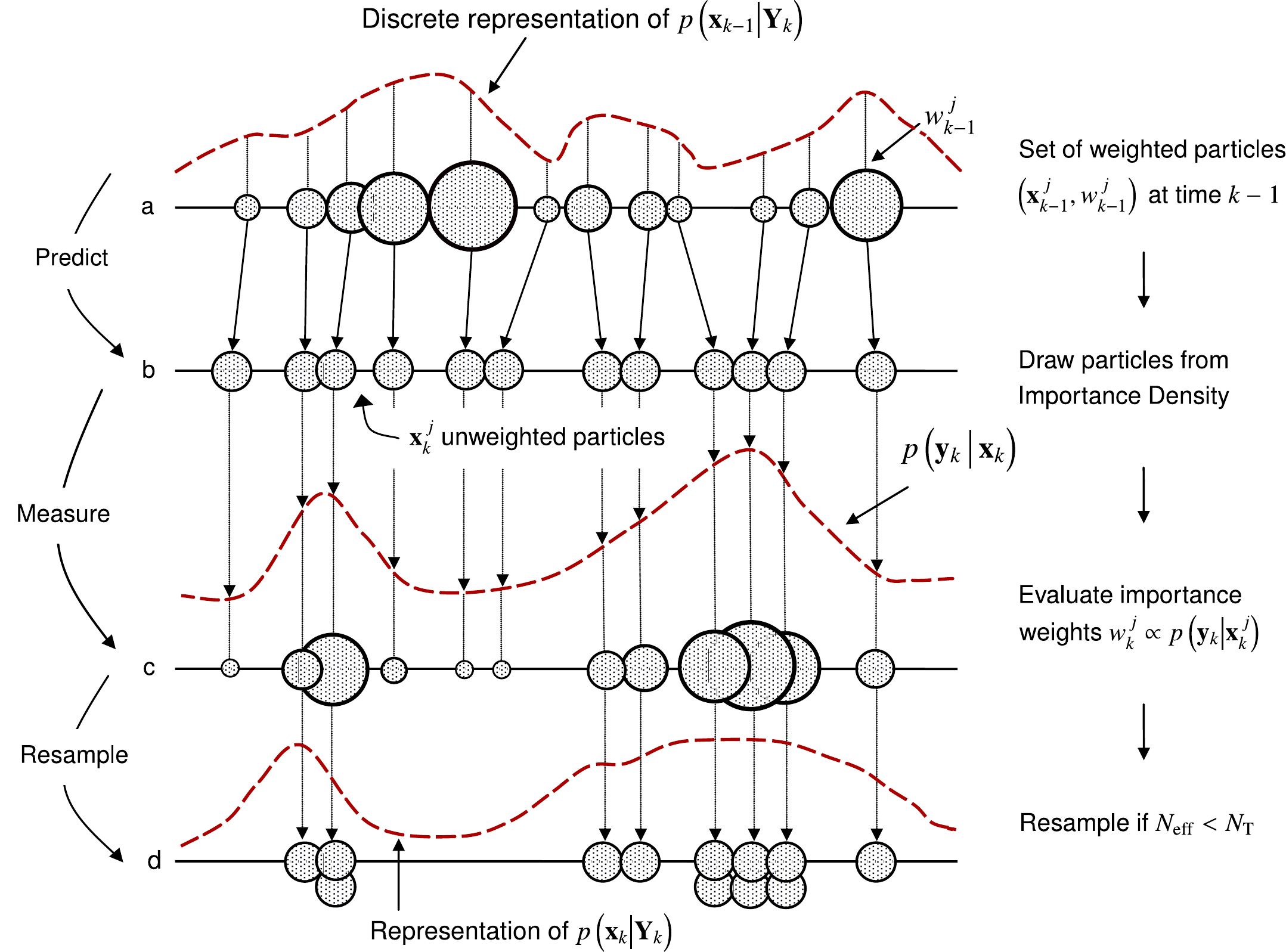}
		\caption{Schematic representation of the standard Particle Filter (PF) algorithm: a) Predict by drawing samples from the importance density; b) Evaluate the importance weights based on the likelihood function; c) Resample if the effective number of particles is below $N_{\mathrm{T}}$ and normalize weights; d) Approximate the posterior density through the set of weighted particles.}
		\label{fig:PFfig}
	\end{center}
\end{figure}

The convergence of all Monte Carlo methods, including PFs, is guaranteed by the central limit theorem as the number of particles approaches infinity however, an increased number of particles might lead to a significant computational cost which can be a major disadvantage of the method. On the other hand, although the UKF is characterized by more limited applicability, since it contains the assumption of Gaussian approximations, it is in general considerably faster the PFs, 
\color{edits}
which is owed to the limited and predefined number of sample points.
\color{black}

\begin{algorithm}[t!]
	\small
	\SetAlgoLined
	\DontPrintSemicolon
	\SetKwInOut{Input}{Input}
	\SetKwInOut{Output}{Output}
	\SetKw{And}{and}
	
	\BlankLine
	Initialize by drawing samples $\mathbf{x}_0^{\,j} \sim p\left(\mathbf{x}_0\right)$ for $j=1,2,\ldots,N$\;
	
	\BlankLine
	\For{$k = 1,\, 2,\, \ldots\, \,$}{
		
		\BlankLine
		(a) Importance sampling\;
		\BlankLine
		
		\For{$j = 1,\, 2,\, \ldots,\, N$}{
			Draw samples $\mathbf{x}_k^{\,j}$ $\sim$ $q\left(\mathbf{x}_k \big| \mathbf{x}_{k-1}^{\,j}, \mathbf{Y}_k \right)$ \;
			Evaluate the importance weights $w_{k}^{\,j} = w_{k-1}^{\,j}\,p\left(\mathbf{y}_k\big|\mathbf{x}_k^{\,j} \right)$\;
		}
		For $j = 1,\, 2,\, \ldots,\, N$ normalize weights $w_k^{\,j} \leftarrow w_k^{\,j}\left[\sum_{j=1}^N w_k^{\,j} \right]^{-1}$\;
		
		\BlankLine
		(b) Resampling
		\BlankLine
		
		Multiply or supress particles $\mathbf{x}_k^{\,j}$  to obtain samples distributed according to $p\left(\mathbf{x}_k\big|\mathbf{Y}_k \right)$\;\vspace{1mm}
		
		For $j = 1,\, 2,\, \ldots,\, N$ reset weights $w_k^{\,j} = 1/N$\;
		
		\BlankLine
		(c) Inference
		
		$\mathbb{E}\left[\mathbf{g}\left(\mathbf{x}_k\right) \big| \mathbf{Y}_k \right] \approx \sum\limits_{j=1}^N w_k^{\,j}\mathbf{g}\left(\mathbf{x}_k^{\,j}\right)$
	}
	
	\caption{Standard Particle Filter (PF)}
	\label{alg:PF}
\end{algorithm}

\subsubsection{Resampling and Sample Impoverishment}
\label{subsubsec:re-sampling}

A standard issue associated with PFs is the problem of degeneracy, which refers to the tendency of a single particle to dominate the weights as the evaluation steps progress. This is a known pathology, which is owed to the inherently increasing variance of the importance weights over time \cite{Doucet2000a}. However, it implies that the bulk of computational effort is wasted for the propagation of non-contributing particles. The effective sample size \cite{Liu1996} is a representative metric of the degeneracy level, which can be estimated as
\begin{equation}\label{eq:Neff}
	N_{\mathrm{ef\-f}}=\frac{1}{\sum\limits_{j=1}^{N}{\left(w_k^{\,j}\right)^2}}
\end{equation}
Sample degeneracy may be tackled either by increasing the number of employed particles or, more effectively, by resampling the particles upon exceedance of a certain degeneracy threshold $N_{\mathrm{T}}$, which is a user-specified variable. The latter technique aims at discarding the particles with negligible weight and enhancing the ones with larger weights. As such, a new set of particles $\left\{\mathbf{x}_k^{j*},\quad  j=1,..,N\right\}$ is generated, which occurs by replacement from the original set, so that $p\left(\mathbf{x}_k^{j*}=\mathbf{x}_k^{\,j}\right) = w_k^{\,j}$. The process is schematically shown in \cref{fig:resample}, where it should be further noted that all weights are reset to $w_k^{\,j} = 1/N$  upon resampling.

\begin{figure}[b!]
	\begin{center}
		\includegraphics[width=0.8\columnwidth]{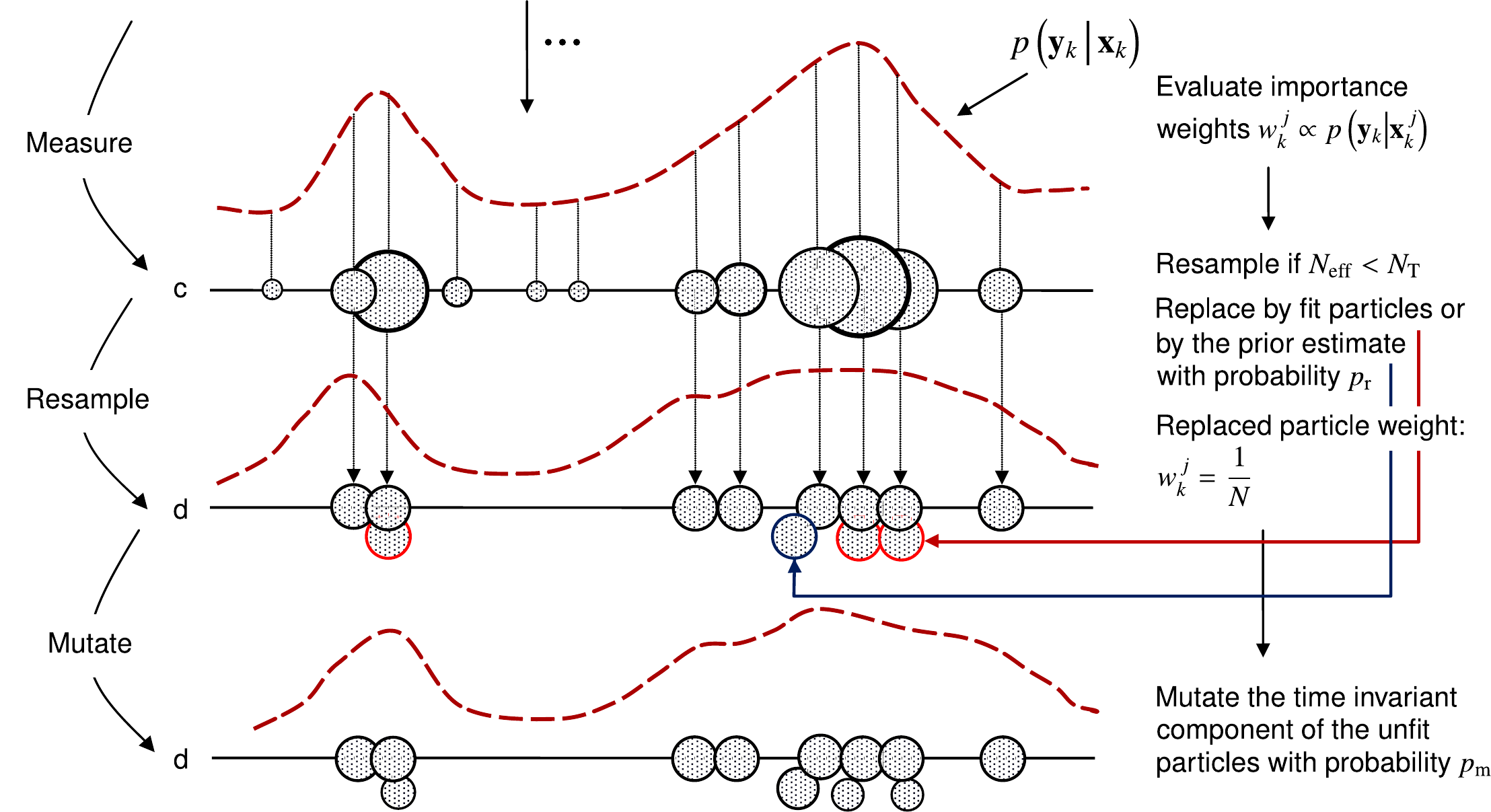}
		\caption{Schematic representation of the Particle Filter with Mutation (MPF): b) Evaluate the importance weights based on the likelihood function; c) Resample if the effective number of particles is below $N_{\mathrm{T}}$ (replace by the fit particles or the prior estimate of the state) and normalize weights; d) Mutate the resampled particles; e) Approximate the posterior density through the set of weighted particles.}
		\label{fig:MPFfig}
	\end{center}
\end{figure}

Despite the efficient treatment of degeneracy, resampling may lead to the so-called sample impoverishment \cite{Arulampalan2002}. This issue may typically arise in problems with small process noise and subsequently result to large-weight particles being selected multiple times, destroying thus the diversity among particles. This problem has been addressed in the literature using evolutionary operations \cite{Park2007} and Support Vector Regression (SVR) \cite{Zhu2005}, as well as combination of PFs with the Kalman filter for the time update step and importance density generation \cite{VanderMerwe2004}. The standard PF implementation, which is described in this section, is schematically presented in \cref{fig:PFfig}, while the detailed steps of the algorithm are listed in Algorithm \ref{alg:PF}.

\subsubsection{The Particle Filter with Mutation (MPF)}
\label{subsec:MPF}

An improvement to the standard PF, in terms of the sample impoverishment problem, is presented in \cite{Chatzi2013} where a mutation operation is introduced for the time invariant components of the state vector, which is used to modify the resampling step. In the standard PF implementation, the unfit particles are initially replaced by fitter ones that are already contained in the pool and then uniform weights are assigned to all particles. Such a strategy may lead to a fast convergence of the time invariant parameters to one of the initial particles, especially in the case of low process noise, as also underlined also in \cite{Chatzi2009}. In this sense, the search space of the state parameters in the standard PF is largely limited by the initially proposed density and particles. This implies that for a successful estimation, the particles should be drawn from the vicinity of the actual parameter values. 
\color{edits}
It should be noted that the importance sampling scheme of the MPF is identical to the one used for the PF and the mutation operation is intended to alleviate the sample impoverishment problem by maintaining diversity in the particles, similarly to other resampling alternatives, such as the Markov Chain Monte Carlo (MCMC) scheme \cite{Gilks2001}.
\color{black}

\begin{algorithm}[h]
	\small
	\SetAlgoLined
	\DontPrintSemicolon
	\SetKwInOut{Input}{Input}
	\SetKwInOut{Output}{Output}
	\SetKw{And}{and}
	
	\BlankLine
	Initialize by drawing samples $\mathbf{x}_0^{\,j} \sim p\left(\mathbf{x}_0\right)$ for $j=1,\, 2,\, \ldots,\, N$\;
	
	\BlankLine
	\For{$k = 1,\, 2,\, \ldots \, $}{
		
		\BlankLine
		(a) Importance sampling\;
		\BlankLine
		
		\For{$j = 1,\, 2,\, \ldots,\, N \, $}{
			Draw samples $\mathbf{x}_k^{\,j}$ $\sim$ $q\left(\mathbf{x}_k \big| \mathbf{x}_{k-1}^{\,j}, \mathbf{Y}_k \right)$ \;\vspace{1mm}
			Evaluate the importance weights $w_{k}^{\,j} = w_{k-1}^{\,j}\,p\left(\mathbf{y}_k\big|\mathbf{x}_k^{\,j} \right)$\;
		}
		For $j = 1,\, 2,\, \ldots,\, N$ normalize weights $w_k^{\,j} \leftarrow w_k^{\,j}\left[\sum_{i=1}^N w_k^{\,i} \right]^{-1}$\;
		
		\BlankLine
		(b) Resampling
		\BlankLine
		
		Multiply or supress particles $\mathbf{x}_k^{\,j}$  to obtain samples distributed according to $p\left(\mathbf{x}_k\big|\mathbf{Y}_k \right)$\;\vspace{1mm}
		
		\For{$j = 1,\, 2,\, \ldots,\, N_{\mathrm{u}}\, $}{
			Replace particle $\mathbf{x}_k^{\,j}$ with probabilty $p_{\mathrm{r}}$ by the prior estimate $\mathbf{x}_{k|k-1}=\mathbf{G}\left(\hat{\mathbf{x}}_{k-1|k-1}, \mathbf{p}_k \right)$ \;\vspace{1mm}
			Mutate the time-invariant components of $\mathbf{x}_k^{\,j}$ with probability $p_{\mathrm{m}}$ using \cref{eq:mutation}\;\vspace{2mm}
			Adjust the weights of mutated particles using \cref{eq:mutation-weights}\;
		}
		
		For $j = 1,\, 2,\, \ldots,\, N$ normalize weights $w_k^{\,j} \leftarrow w_k^{\,j}\left[\sum_{i=1}^N w_k^{\,i} \right]^{-1}$\;
		
		\BlankLine
		(c) Inference
		
		$\mathbb{E}\left[\mathbf{g}\left(\mathbf{x}_k\right) \big| \mathbf{Y}_k \right] \approx \sum\limits_{j=1}^N w_k^{\,j}\mathbf{g}\left(\mathbf{x}_k^{\,j}\right)$
	}
	
	\caption{Particle Filter with Mutation (MPF)}
	\label{alg:MPF}
\end{algorithm}

In the MPF algorithm, the $N_{\mathrm{u}}$ unfit particles are replaced, with a user-defined probability $p_{\mathrm{r}}$, by the prior state estimate $\hat{\mathbf{x}}_{k|k-1} = \mathbf{G}\left(\hat{\mathbf{x}}_{k-1|k-1}, \mathbf{p}_{k-1} \right)$ and the time invariant components of the state vector are mutated in order to allow the exploration of the parameter space. Concretely, the mutation is applied to each $i$-th parameter component of the $j$-th particle with a probability $p_m$, using the following perturbation rule

\begin{equation}
	\mathbf{x}_k^{\,i,j} = \mathbf{x}_k^{\,i,j}\cdot\left(1+ d_i\cdot(m_i-0.5) \right)
	\label{eq:mutation}
\end{equation}

in which $d_i$ denotes the perturbation radius and $m_i\sim\mathcal{U}\left(0,1 \right)$ is a random number drawn from the standard uniform distribution. This operation resembles the creep mutation of the traditional Genetic Algorithm (GA) which takes place in the phenotype, i.e. the real representation of the parameter strings \cite{Whitley1994}. Thereafter, the mutated particles are assigned a weight which is inversely proportional to the relative difference $\Delta\mathbf{x}_{k}^{\,j}$ between the mutated and the parent particle, calculated as follows
\begin{equation}
	w_k^{\,j} = 
	\frac{1}{N}\frac{1}{\dfrac{\left\|\Delta \mathbf{x}_k^{\,j}\right\|}{\left\|\mathbf{x}_k^{\,}\right\|}+1}
	\label{eq:mutation-weights}
\end{equation}
where $\left\| \bullet \right\|$ denotes the $L_2$ norm, 
\color{edits}
although an alternate penalty weighting scheme of similar logic may be readily adopted. 
\color{black}
On the other hand, the weights of the non-mutated re-sampled particles remain unchanged, equal to $w_k^{\,j} = 1/N$, and the entire set of weights in subsequently normalized. \cref{eq:mutation-weights} offers a heuristic approach to specifying a fast computed term term that penalizes heavily mutated particles; this could alternatively by replaced by an alternate distance metric, which includes consideration of the likelihood. A graphical representation of the MPF is shown in \cref{fig:MPFfig}, while the detailed steps of the algorithm are listed in Algorithm \ref{alg:MPF}.

\begin{algorithm}[h]
	\small
	\SetAlgoLined
	\DontPrintSemicolon
	\SetKwInOut{Input}{Input}
	\SetKwInOut{Output}{Output}
	\SetKw{And}{and}
	
	\BlankLine
	Initialize by drawing samples $\mathbf{x}_0^{\,j} \sim p\left(\mathbf{x}_0\right)$ for $j=1,\, 2,\, \ldots,\, N$\;
	
	\BlankLine
	\For{$k = 1,\, 2,\, \ldots \,$}{
		
		\BlankLine
		(a) Importance sampling\;
		\BlankLine
		
		\For{$j = 1,\, 2,\, \ldots,\, N\,$}{
			
			Calculate sigma points $\mathbfcal{X}_{k-1|k-1}^{\,j}$ for particle $\mathbf{x}_{k-1}^{\,j}$ using \cref{eq:sigma-points} \; \vspace{1mm}
			
			Perform a time update step to obtain $\hat{\mathbf{x}}_{k|k-1}^{\,j}$, $\mathbf{P}_{k|k-1}^{\,j}$ using \cref{eq:UKF-prediction-1a,eq:UKF-prediction-1b} \; \vspace{1.5mm}
			
			Perform a measurement update step to obtain $\hat{\mathbf{x}}_{k|k}^{\,j}$, $\mathbf{P}_{k|k}^{\,j}$ using \cref{eq:UKF-update-1a,eq:UKF-update-1b} \; \vspace{1mm}
			
			Draw samples $\mathbf{x}_k^{\,j}$ $\sim$ $q\left(\mathbf{x}_k \big| \mathbf{x}_{k-1}^{\,j}, \mathbf{Y}_k \right) \approx \mathcal{N}\left(\mathbf{x}_k; \hat{\mathbf{x}}_{k|k}, \mathbf{P}_{k|k} \right)$ \; \vspace{0.5mm}
			Evaluate the importance weights $w_{k}^{\,j} = w_{k-1}^{\,j}\,p\left(\mathbf{y}_k\big|\mathbf{x}_k^{\,j} \right)$\;
		}
		For $j=1,\, 2,\, \ldots,\, N$ normalize weights $w_k^{\,j} \leftarrow w_k^{\,j}\left[\sum_{j=1}^N w_k^{\,j} \right]^{-1}$\;
		
		\BlankLine
		(b) Resampling
		\BlankLine
		
		Multiply or supress particles $\mathbf{x}_k^{\,j}$  to obtain samples distributed according to $p\left(\mathbf{x}_k\big|\mathbf{Y}_k \right)$\;\vspace{1mm}
		
		For $j = 1,\, 2,\, \ldots,\, N$ reset weights $w_k^{\,j} = 1/N$ \;
		
		\BlankLine
		(c) Inference
		
		$\mathbb{E}\left[\mathbf{g}\left(\mathbf{x}_k\right) \big| \mathbf{Y}_k \right] \approx \sum\limits_{j=1}^N w_k^{\,j}\mathbf{g}\left(\mathbf{x}_k^{\,j}\right)$
	}
	
	\caption{Sigma-Point Particle Filter (SPPF)}
	\label{alg:SPPF}
\end{algorithm}


\subsection{Sigma-Point Particle Filter (SPPF)}
\label{subsec:SPPF}

\begin{figure}[h]
	\centering
	\includegraphics[width=0.65\textwidth]{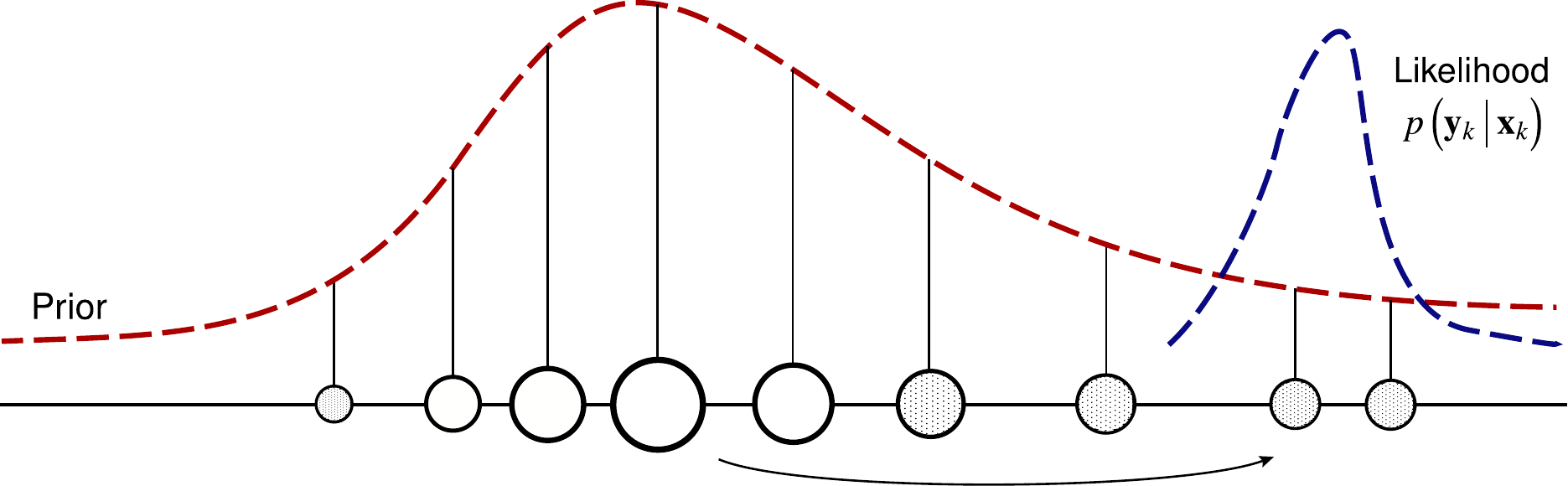}
	\caption{Schematic representation of the particle movement towards the likelihood region}
	\label{fig:sppf}
\end{figure}

An alternative improvement to the standard PF in terms of the proposal distribution is provided by the Sigma-Point Particle Filter (SPPF). This algorithm has been proposed in \cite{VanderMerwe2004}, on the basis of the work published in \cite{Ito2000} on Gaussian filters, whereby the particles are moved towards the areas of high likelihood, as defined by the current observation $\mathbf{y}_k$. The idea of this algorithm, which is schematically presented in \cref{fig:sppf} and is intended to address the impoverishment issue, is materialized by assigning a Gaussian proposal distribution $\mathcal{N}\left(\mathbf{x}_{k};\mathbf{x}_k^{\,j},\mathbf{P}_k^{\,j} \right)$ to each particle $\mathbf{x}_k^{\,j}$, in order to calculate the mean and covariance of the importance distribution. These can be obtained using an EKF, or more efficiently a UKF, which avoids the linearization of system equations and additionally accounts for higher order statistics. Thereafter, each one of the Gaussian proposal distributions $\mathcal{N}\left(\mathbf{x}_{k};\mathbf{x}_k^{\,j},\mathbf{P}_k^{\,j} \right)$ for $j=1,2,\ldots,N$ is used to draw the $j$-th particle at time $k$.

The detailed steps of the SPPF algorithm are listed in Algorithm \ref{alg:SPPF}. These are herein derived on the basis of the UKF, however, these can be substituted by an EKF or even by a Square-Root Central Difference Kalman filter (SR-CDKF), as proposed in \cite{VanderMerwe2004}. It should be also noted that despite the fact that the SPPF is based on the Gaussian assumption, it has been shown to provide a better approximation to the target distribution for a number of applications \cite{Freitas1999}.

\subsection{Gaussian Mixture Sigma-Point Particle Filter (GMSPPF)}
\label{subsec:GMSPPF}

In terms of performance, the SPPF has been shown to address the sample depletion problem and outperform the standard PF in terms of accuracy, however, it is still characterized by a considerable computational cost. This is owed to the fact that each particle is associated with a UKF, or alternatively an EKF, which in turn requires a system linearization for each particle.

\begin{algorithm}[b!]
	\small
	\SetAlgoLined
	\DontPrintSemicolon
	\SetKwInOut{Input}{Input}
	\SetKwInOut{Output}{Output}
	\SetKw{And}{and}
	
	\BlankLine
	Specify the number of mixing components and initialize by drawing samples $\mathbf{x}_0^{\,j} \sim p\left(\mathbf{x}_0\right)$ for $j=1,2,\ldots,N$\;\vspace{1mm}
	
	\For{$k = 1,\, 2,\, \ldots \,$}{
		
		\BlankLine
		(a) Time update\;
		\BlankLine
		
		Set $p_q\left(\mathbf{v}_{k-1} \right) = \mathcal{N}\left(\mathbf{v}_{k-1};\bar{\mathbf{v}}_{k-1}^{\,q},\mathbf{Q}_{k-1}^{\,q} \right)$ for $q=1,2,\ldots, G_{\mathrm{p}}$\;\vspace{0.5mm}
		
		Set $p_s\left(\mathbf{x}_{k-1}\big|\mathbf{Y}_{k-1} \right) = \mathcal{N}\left(\mathbf{x}_{k-1};\hat{\mathbf{x}}_{k-1|k-1}^{\,s},\mathbf{P}_{k-1|k-1}^{\,s} \right)$ for $s=1,2,\ldots, G_{\mathrm{s}}$\;\vspace{0.5mm}
		
		Set $p\left(\mathbf{w}_{k}\right) =  \mathcal{N}\left(\mathbf{w}_{k};\bar{\mathbf{w}}_{k}^{\,r},\mathbf{R}_{k}^{\,r} \right)$ for $r=1,2,\ldots, G_{\mathrm{m}}$\;\vspace{0.5mm}
		
		
		\For{$s^{\,-}=(s,q)$ with $s\in\big\{1,\, 2,\, \ldots,\, G_{\mathrm{s}}\, \Big\}$ \text{and} $q\in\Big\{1,\, 2,\, \ldots,\, G_{\mathrm{p}}\Big\}$}{
			Calculate $p_{s^{\,-}}\left(\mathbf{x}_k\big|\mathbf{Y}_{k-1} \right) = \mathcal{N}\left(\mathbf{x}_k;\hat{\mathbf{x}}_{k|k-1}^{\,s\,-},\mathbf{P}_{k|k-1}^{\,s\,-} \right)$ and update the coefficients $\alpha_k^{\,\mathrm{s}\,-} =
			\alpha_{k-1}^{\,s}\, \beta_{k-1}^{\,q}\, \Big/ \, \left( \sum\limits_{s=1}^{G_{\,\mathrm{s}}} \sum\limits_{q=1}^{G_{\,\mathrm{p}}} \alpha_{k-1}^{\,s}\,\beta_{k-1}^{\,q} \right)$
		}
		
		\For{$s^{\,+}=(s^{\,-},r)$ with $s^{\,-}\in\left\{1,\, 2,\, \ldots,\, G_{\mathrm{s}}^{\,-}\right\}$ and $r\in\left\{1,\, 2,\, \ldots,\, G_{\mathrm{m}}\right\}$}{\vspace{1mm}
			Compute the likelihood $s_{k}^{\,r} = p_r\left(\mathbf{y}_k|\mathbf{x}_{k} = \hat{\mathbf{x}}_{k|k-1}^{\,s\,-} \right)$\;\vspace{-0.5mm}
			Calculate $p_{s^{\,+}}\left(\mathbf{x}_k\big|\mathbf{Y}_k \right) = \mathcal{N}\left(\mathbf{x}_k;\hat{\mathbf{x}}_{k|k}^{\,s\,+},\mathbf{P}_{k|k}^{\,s\,+} \right)$ and update
			$\alpha_k^{\,\mathrm{s}\,+} =
			\alpha_{k}^{\,s\,-}\, \gamma_{k}^{\,r} s_{k}^{\,r}\, \Big/ \,\left( \sum\limits_{s^{\,-}=1}^{G_{\,\mathrm{s}}^{\,-} } \sum\limits_{r=1}^{G_{\,\mathrm{r}}} \alpha_{k}^{\,s\,-}\,\gamma_{k}^{\,r}s_{k}^{\,r} \right)$
		}
		
		Calculate the prior state density $p\left(\mathbf{x}_k\big|\mathbf{Y}_{k-1}\right)$ using \cref{eq:GMMa}\;\vspace{1mm}
		Calculate the posterior state density $p\left(\mathbf{x}_k\big|\mathbf{Y}_k\right)$ using \cref{eq:GMMb}\;
		
		\BlankLine
		(b) Measurement update\;
		\BlankLine
		
		Draw particles $\mathbf{x}_k^{\,j}$ from the posterior density $p\left(\mathbf{x}_k\big|\mathbf{Y}_k\right)$ obtained in step 16\; \vspace{1mm}
		
		For $j=1,\, 2,\, \ldots,\, N$ calculate the particle weights $w^{\,j} = p\left(\mathbf{y}_k \big| \mathbf{x}_k^{\,j} \right)p\left(\mathbf{x}_{\,k}^j\big|\mathbf{Y}_{k-1} \right)\Big/p\left(\mathbf{x}_{\,k}^j\big|\mathbf{Y}_{k} \right)$\;\vspace{0.5mm}
		
		For $j=1,\, 2,\, \ldots,\, N$ normalize weights $w_k^{\,j} \leftarrow w_k^{\,j}\left[\sum_{i=1}^N w_k^{\,i} \right]^{-1}$ \; \vspace{1mm}
		
		Fit the $G_{\mathrm{s}}$-component GMM of \cref{eq:GMM-1} to the particles $\mathbf{x}_{k}^{\,j}$ drawn in step 18 \;
		
		\BlankLine
		(c) Inference
		
		$\mathbb{E}\left[\mathbf{g}\left(\mathbf{x}_k\right)\big|\mathbf{Y}_k \right] = \sum\limits_{s=1}^{G_{\mathrm{s}}}\alpha_k^{\,s}\,\mathbb{E}\left[\mathbf{g}\left(\mathbf{x}_k^{\,s}\right)\big|\mathbf{Y}_k \right]$
		
	}
	
	\caption{Gaussian Mixture Sigma-point Particle Filter (GMSPPF)}
	\label{alg:GMSPPF}
\end{algorithm}

A further improvement to the shortcomings of the computational performance of the SPPF has been achieved by the Gaussian Mixture Sigma-Point Particle Filter (GMSPPF)
\color{edits}
, which has been proposed in \cite{VanderMerwe2004} and is especially suited for simulation of problems with multi-modal probability density functions.
\color{black}
This algorithm is based on i) an importance sampling step for the measurement update and ii) a UKF-based Gaussian sum filter, which is used for the time update step and the generation of the proposal density. As such, a $G_{\mathrm{s}}$-component Gaussian Mixture Model (GMM) is used at each time step for the approximation of the posterior state density $p\left(\mathbf{x}_{k-1}\big|\mathbf{Y}_{k-1} \right)$, as follows
\begin{equation}
	p\left(\mathbf{x}_{k-1}\big|\mathbf{Y}_{k-1} \right) = 
	\sum_{s=1}^{G_{\mathrm{s}}} \alpha_{k-1}^{\,s}\mathcal{N}\left(\mathbf{x}_{k-1}; \hat{\mathbf{x}}^{\,s}_{k-1|k-1},\mathbf{P}^{\,s}_{k-1|k-1} \right)
	\label{eq:GMM-1}
\end{equation}
where $\alpha_{k-1}^{\,s}$ is the weight of the $s$-th component and $\mathcal{N}\left( \mathbf{x}; \mathbf{m},\mathbf{P} \right)$ represents the normal distribution of the $s$-th component with argument $\mathbf{x}$, mean value $\mathbf{m}$ and covariance matrix $\mathbf{P}$. The density of \cref{eq:GMM-1} is obtained from the posterior particles extracted after the measurement update step. This calculation is performed with the use of the Expectation-Maximization (EM) algorithm and it further implies that not only the state densities but also the ones for the noise terms, $p(\mathbf{v}_{k-1})$ and $p(\mathbf{w}_k)$, are also represented by GMMs, whose coefficients are $\beta_{k-1}^{\,q}$ for $q=1,2,\ldots,G_{\mathrm{p}}$ and $\gamma_{k}^{\,r}$ for $r=1,2,\ldots,G_{\mathrm{m}}$, respectively. As such, prior and posterior densities, which are obtained from the prediction and measurement steps, are also expressed by the following GMMs
\begin{align}
	p\left(\mathbf{x}_{k}\big|\mathbf{Y}_{k-1} \right) & = 
	\sum_{s^{\,-}=1}^{G_{\mathrm{s}}^{\,-}}\alpha_{k}^{\,s\,-}\mathcal{N}\left(\mathbf{x}_{k}; \hat{\mathbf{x}}^{\,s\,-}_{k|k-1},\mathbf{P}^{\,s\,-}_{k|k-1}  \right)
	\label{eq:GMMa}\\
	p\left(\mathbf{x}_{k}\big|\mathbf{Y}_{k} \right) & = 
	\sum_{s^{\,+}=1}^{G_{\mathrm{s}}^{\,+}}\alpha_{k}^{\,s\,+}\mathcal{N}\left(\mathbf{x}_{k}; \hat{\mathbf{x}}^{\,s\,-}_{k|k},\mathbf{P}^{\,s\,-}_{k|k}\right)
	\label{eq:GMMb}
\end{align}
where $G_{\mathrm{s}}^{\,-} = G_{\mathrm{s}}\cdot G_{\mathrm{p}}$ and $G_{\mathrm{s}}^{\,+} = G_{\mathrm{s}}^{\,-}\cdot G_{\mathrm{m}}$. From the above equations it is suggested that the number of mixing components increases in each prediction and correction step. Namely, a $G_{\mathrm{s}}$-component GMM is initially used for the target density, which increases to $G_{\mathrm{s}}^{\,-}$ in the prediction step and subsequently to $G_{\mathrm{s}}^{\,+}$ in the correction step. This implies an exponential increase in the number of mixing terms, which can be avoided by fitting a $G_{\mathrm{s}}$-component GMM to the $N$ particles drawn from the $G_{\mathrm{s}}^+$-component GMM of the posterior density, whose proposal is obtained from \cref{eq:GMMb}. Lastly, the inference step is carried out with the use of the fitted GMM, as documented in Algorithm \ref{alg:GMSPPF}, which contains all the steps of the GMSPPF.

%


\subsection{Rao-Blackwellised Particle Filter (RBPF)}
\label{subsec:RBPF}

In certain problems the model dynamics may contain a tractable subspace, which can be marginalised, as is the case for problems of joint state-parameter estimation, where the time invariant parameters evolve differently to the system's dynamical states. Marginalisation results in a more efficient PF implementation, 
\color{edits}
the so called Rao-Blackwellised Particle Filter (RBPF) \cite{Li2003,Goodall2005},
\color{black}
which allows the analytical computation of the filtering equations related to the marginalized state. To do so, the state vector is partitioned as follows
\begin{equation}
	\mathbf{x}_k = 
	\begin{bmatrix}
		\mathbf{x}_k^{\,\mathrm{a}}\\[2mm]
		\mathbf{x}_k^{\,\mathrm{b}}\\[1mm]
	\end{bmatrix}
\end{equation}
where $\mathbf{x}_k^{\,\mathrm{b}}$ is the part of the state vector to be marginalized and $\mathbf{x}_k^{\,\mathrm{a}}$ contains the remaining state vector. Then using Bayes rule, the posterior PDF can be rewritten as

\begin{equation}
	p\left(\mathbf{x}_k\big| \mathbf{Y}_k \right) = 
	p\left(\mathbf{x}_k^{\,\mathrm{a}},\mathbf{x}_k^{\,\mathrm{b}}\,\big| \mathbf{Y}_k \right) = p\left(\mathbf{x}_k^{\,\mathrm{b}}\,\big|\,\mathbf{x}_k^{\,\mathrm{a}},\mathbf{Y}_k \right)p\left(\mathbf{x}_k^{\,\mathrm{a}}\,\big| \mathbf{Y}_k \right)
	\label{eq:rbpf-decomposition}
\end{equation}

and the expectation over the density $p\left(\mathbf{x}_k\big| \mathbf{Y}_k \right)$ can be also decomposed as follows

\begin{align}
	\mathbb{E}\left[\mathbf{g}\left(\mathbf{x}_{k}\right)\big|\mathbf{Y}_{k}\right] & = \int \mathbf{g}\left(\mathbf{x}_{k}\right)\, p\left(\mathbf{x}_k\big|\mathbf{Y}_{k}\right)\,d\mathbf{x}_k\\[1mm]
	& = \int\left[\int\mathbf{g}\left(\mathbf{x}_k^{\,\mathrm{a}},\mathbf{x}_k^{\,\mathrm{b}} \right)p\left(\mathbf{x}_k^{\,\mathrm{b}}\,\big|\,\mathbf{x}_k^{\,\mathrm{a}},\mathbf{Y}_k\right)d\mathbf{x}_k^{\,\mathrm{b}}\right]p\left(\mathbf{x}_k^{\,\mathrm{a}}\,\big|\,\mathbf{Y}_k\right)d\mathbf{x}_k^{\,\mathrm{a}}
\end{align}

which implies that if the integral in the brackets can be evaluated analytically, the PF need be used only for the sample-based approximation of $p\left(\mathbf{x}_k^{\,\mathrm{a}}\,\big|\,\mathbf{Y}_k \right)$. Therefore, the above expectation can be rewritten as 

\begin{equation}
	\mathbb{E}\left[\mathbf{g}\left(\mathbf{x}_{k}\right)\big|\mathbf{Y}_{k}\right]
	= \sum_{j=1}^{N}w_k^{\,j}\int\mathbf{g}\left(\mathbf{x}_k^{\,\mathrm{a}},\mathbf{x}_k^{\,\mathrm{b}} \right)p\left(\mathbf{x}_k^{\,\mathrm{b}}\,\big|\,\mathbf{x}_k^{\,\mathrm{a}},\mathbf{Y}_k\right)d\mathbf{x}_k^{\,\mathrm{b}}
	\label{eq:rbpf-inference}
\end{equation}

where $w_k^{\,j}$ for $j=1,2,\ldots,N$ are the importance weights associated with the particles of $\mathbf{x}_k^{\,\mathrm{a}}$. As such, the reduced in size $\mathbf{x}_k^{\,\mathrm{a}}$ will require less particles than the full size state $\mathbf{x}_k$, increasing thus the accuracy of estimates for a given amount of computational resources. In the context of filtering, such an analytical computation is carried using a Kalman filter, implying thus that $\mathbf{x}_k^{\,\mathrm{b}}$ is described by a linear transition model and Gaussian noise terms, which is typically the case in parameter estimation problems where a random walk model with Gaussian noise is assumed for the parameter evolution. Despite this assumption, the RBPF algorithm is herein presented for the general case, in which the transition model of $\mathbf{x}_{k}^{\,\mathrm{b}}$ may be described by any distribution $p\left(\mathbf{x}_k^{\,\mathrm{b}}\,\big|\,\mathbf{x}_k^{\,\mathrm{a}},\mathbf{Y}_k \right)$. 

\begin{algorithm}[h]
	\small
	\SetAlgoLined
	\DontPrintSemicolon
	\SetKwInOut{Input}{Input}
	\SetKwInOut{Output}{Output}
	\SetKw{And}{and}
	
	Initialize the state $\mathbf{x}_{\,0}^{\,\mathrm{a}}$ by drawing samples $\mathbf{x}_{\,0}^{\,\mathrm{a},j}\sim p\left(\mathbf{x}_{\,0}^{\,\mathrm{a}} \right)$ for $j=1,\, 2,\, \ldots,\, N$\;\vspace{1mm}
	
	Initialize the $j$-th filter associated with $\mathbf{x}_{\,0}^{\,\mathrm{a},j}$ by specifying $\hat{\mathbf{x}}_{\,0|0}^{\,\mathrm{b},j}$ and $\mathbf{P}_{\,0|0}^{\,\mathrm{b},j}$ for $j=1,\, 2,\, \ldots,\, N$\;\vspace{1mm}
	
	\For{$k = 1,\, 2,\, \ldots\, $}{
		\BlankLine
		(a) Time update\; \vspace{1mm}
		
		\For{$j = 1,\, 2\, \ldots,\, N$}{
			Perform a time update step to obtain the prior state statistics $\hat{\mathbf{x}}_{k|k-1}^{\,\mathrm{b},j}$ and $\mathbf{P}_{k|k-1}^{\,\mathrm{b},j}$\; \vspace{1mm}
			
			Draw samples $\mathbf{x}_k^{\,\mathrm{a},j} \sim q\left(\mathbf{x}_k^{\,\mathrm{a}}\big|\mathbf{x}_{k-1}^{\,\mathrm{a},j},\mathbf{Y}_k \right)$\;\vspace{1mm}
			
			Calculate the particle weights using \cref{eq:rbpf-weights}\;
		}
		For $j=1,\, 2,\, \ldots,\, N$ normalize weights $w_k^{\,j} \leftarrow w_k^{\,j}\left[\sum_{j=1}^N w_k^{\,j} \right]^{-1}$ \; 
		
		\BlankLine
		(b) Measurement update\;\vspace{1mm}
		
		\For{$j = 1,\, 2,\, \ldots, \,N$}{
			Perform a measurement update step to obtain the posterior state statistics $\hat{\mathbf{x}}_{k|k}^{\,\mathrm{b},j}$ and $\mathbf{P}_{k|k}^{\,\mathrm{b},j}$ conditional on $\mathbf{x}_{k}^{\,\mathrm{a},j}$\;
		}
		
		\BlankLine
		(c) Resampling\;
		\BlankLine
		
		Multiply or supress particles $\mathbf{x}_k^{\,\mathrm{a},j}$  to obtain samples distributed according to $p\left(\mathbf{x}_k^{\,\mathrm{a}}\big|\mathbf{Y}_k \right)$\;\vspace{1mm}
		
		For $j=1,2,\ldots N$ reset weights $w_k^{\,j} = 1/N$\;
		
		\BlankLine
		(d) Inference\;
		
		$\mathbb{E}\left[\mathbf{g}\left(\mathbf{x}_{k}\right)\big|\mathbf{Y}_{k}\right]
		= \mathlarger{\sum}\limits_{j=1}^{N}w_k^{\,j}\mathlarger{\int}		
		\mathbf{g}\left(\mathbf{x}_k^{\,\mathrm{a}},\mathbf{x}_k^{\,\mathrm{b}} \right)p\left(\mathbf{x}_k^{\,\mathrm{b}}\,\big|\,\mathbf{x}_k^{\,\mathrm{a}},\mathbf{Y}_k\right)d\mathbf{x}_k^{\,\mathrm{b}}$\;
	}

	\caption{Rao-Blackwellised Particle Filter (RBPF)}
	\label{alg:RBPF}
\end{algorithm}

Within this context, each particle $\mathbf{x}_k^{\,\mathrm{a},j}$ for $j=1,2,\ldots,N$ that is used for the approximation of $p\left(\mathbf{x}_k^{\,\mathrm{a}}\,\big|\,\mathbf{Y}_k \right)$ is associated with a Bayesian filter for the integration of $p\left(\mathbf{x}_k^{\,\mathrm{b}}\,\big|\,\mathbf{x}_k^{\,\mathrm{a}},\mathbf{Y}_k \right)$, creating thus a hierarchical process in which the expectation over the density $p\left(\mathbf{x}_k\big| \mathbf{Y}_k \right)$ is calculated in two steps, as postulated by \cref{eq:rbpf-decomposition}. Thus, given the particles $\mathbf{x}_{k|k-1}^{\,\mathrm{a},j}$, which are drawn from $p\left(\mathbf{x}_k^{\,\mathrm{a}}\,\big|\,\mathbf{Y}_k \right)$, and the associated prior estimates $\mathbf{x}_{k|k-1}^{\,\mathrm{b},j}$ along with the corresponding covariance matrices $\mathbf{P}_{k|k-1}^{\,\mathrm{b},j}$, which are extracted from the transition model of $\mathbf{x}_{k}^{\,\mathrm{b}}$, one can estimate the importance weights according to

\begin{equation}
	w_k^{\,j} = 	w_{k-1}^{\,j} p\left(\mathbf{y}_k\,\big|\,\mathbf{x}_{k|k-1}^{\,\mathrm{a},j},\mathbf{Y}_{k-1} \right) 
	\label{eq:rbpf-weights}
\end{equation}

which are fully defined by the measurement equation. In this case, it is assumed that the importance density of $\mathbf{x}_k^{\mathrm{a}}$ is identical to the transitional density $p\left(\mathbf{x}_k^{\,\mathrm{a}}\,\big|\,\mathbf{x}_{k-1}^{\,\mathrm{a}} \right)$, resulting thus to \cref{eq:rbpf-weights}. In the more general case, in which a different importance density $q\left(\mathbf{x}_k^{\,\mathrm{a}}\big|\mathbf{x}_{k-1}^{\,\mathrm{a},j},\mathbf{Y}_k \right)$ is assumed, the weights can be calculated from \cref{eq:weights-1}.

It should be noted that when the dynamics of $\mathbf{x}_{k}^{\,\mathrm{b}}$ are governed by a linear Gaussian model, the probability distribution of \cref{eq:rbpf-weights} is a Gaussian one and equal to the marginal measurement likelihood of the corresponding Kalman filter. Thereafter, the prior estimate of $\mathbf{x}_k^{\,\mathrm{a}}$ can be obtained as the following weighted sample mean
\begin{equation}
	\hat{\mathbf{x}}_{k|k-1}^{\,\mathrm{a}} = \sum_{j=1}^N w_k^{\,j}\,\mathbf{x}_{k|k-1}^{\,\mathrm{a},j}
\end{equation}
which should be evaluated upon normalization of the importance weights obtained from \cref{eq:rbpf-weights}. A measurement update step can be performed for each filter associated with $\mathbf{x}_{k|k-1}^{\,\mathrm{a},j}$ for $j=1,2,\ldots,N$ in order to obtain the posterior state estimate $\mathbf{x}_{k|k}^{\,\mathrm{b},j}$ conditional on $\mathbf{x}_{k|k-1}^{\,\mathrm{a},j}$ and subsequently perform the inference step using \cref{eq:rbpf-inference}. The detailed steps for the implementation of the Bayesian inference problem using the RBPF algorithm are listed in Algorithm \ref{alg:RBPF}.

\color{edits}
These steps are formulated for the general case where the prior and posterior distributions of the marginalized state are obtained by any filtering algorithm. In the most simple and computationally efficient variant of the RBPF, which corresponds to linear Gaussian equations of the marginalized state, these distributions are extracted by means of a Kalman filter. However, any of the herein presented algorithms can be used instead, when either of the marginalized state or measurement equations is nonlinear or characterized by non-Gaussian noise. 
In such cases, only a slight computational improvement is offered by the RBPF due to the separation of the sampling space however, the major advantage of the RBPF is still maintained and this pertains, according to the Rao-Blackwell theorem \cite{Berger1985}, to the reduced variance of the estimated quantities.

\color{black}


\section{State, input and parameter estimation}
\label{sec:state-input-parameter}

The filtering algorithms presented in the previous section have been formulated for the state and unknown system parameters estimation. This was accomplished by treating the parameters as random variables, which are appended to the state vector. 
\color{edits}
However, a further typical problem met within the context of SHM is the need for estimating the response, and possibly the unknown system parameters, under additional absence of information in the loads (input). Similarly to what is previously described,
\color{black}
the additional estimation of the unknown excitation can be performed by assuming that each input signal is a random variable, whose dynamics are governed by Gaussian random walk. In this sense, the state-parameter estimation problem, described by \cref{eq:augmented-state,eq:augmented-output}, can be further modified as follows

\begin{align}
	\mathbf{p}_k & \sim \mathcal{N}\left(\mathbf{p}_k;\mathbf{p}_{k-1},\bm{\Sigma}_{k-1}^{\,\mathrm{p}} \right)\\[1mm]
	\tilde{\mathbf{x}}_{k+1} & = \tilde{\mathbf{G}}\left(\tilde{\mathbf{x}}_{k}, \mathbf{p}_k \right)+\tilde{\mathbf{v}}_{k}\\[1mm]
	\mathbf{y}_k & = \tilde{\mathbf{H}}\left(\tilde{\mathbf{x}}_k,\mathbf{p}_k \right)+\mathbf{w}_k
\end{align}

where $\bm{\Sigma}_{k-1}^{\mathrm{p}}$ denotes the covariance matrix of the random walk model assumed for the input. The latter can be also rewritten in the following form
\begin{align}
	\mathbf{p}_{k+1} = \mathbf{p}_k+\mathbf{w}^{\,\mathrm{p}}_k
	\label{eq:input-process}
\end{align}

where $\mathbf{w}_k^{\,\mathrm{p}}$ is the equivalent Gaussian zero-mean input process noise with covariance matrix $\bm{\Sigma}_{k-1}^{\,\mathrm{p}}$.

\begin{algorithm}[t!]
	\small
	
	\caption{Dual Kalman Filter - Unscented Kalman Filter (DKF-UKF)}
	\label{alg:DKF-UKF}
	
	\SetAlgoLined
	\DontPrintSemicolon
	\SetKwInOut{Input}{Input}
	\SetKwInOut{Output}{Output}
	\SetKw{And}{and}
	
	Specify the UKF parameters $\alpha$, $\beta$ and $\kappa$\; \vspace{1mm}
	
	Initialize the state and input distribution parameters $\hat{\mathbf{x}}_{\,0|0}$, $\mathbf{P}_{\,0|0}$ and $\hat{\mathbf{p}}_{\,0|0}$ $\mathbf{P}_{\,0|0}^{\,\mathrm{p}}$ \;\vspace{1mm}
	
	\For{$k = 1,\, 2,\, \ldots\, $}{
		\BlankLine
		
		Calculate sigma points $\mathbfcal{X}_{k-1|k-1}^{j}$ using \cref{eq:sigma-points}\;
		
		\BlankLine
		(a) Time update\; \vspace{1mm}
		\BlankLine
		
		Perform a KF time update step to obtain the prior input estimates $\hat{\mathbf{p}}_{k|k-1}$ and $\mathbf{P}_{k|k-1}^{\mathrm{p}}$\; \vspace{1mm}
		
		Calculate the prior sigma points $\mathbfcal{X}_{k|k-1}^{j}$ by propagating $\mathbfcal{X}_{k-1|k-1}^{j}$ through \cref{eq:state-2}\;\vspace{1mm}
		
		Obtain the prior state statistics $\hat{\mathbf{x}}_{k|k-1}$ and $\mathbf{P}_{k|k-1}$ using \cref{eq:UKF-prediction-1a,eq:UKF-prediction-1b}\;\vspace{1mm}
		
		\BlankLine
		(b) Measurement update step\;\vspace{1mm}
		\BlankLine
		
		Perform a KF measurement update step to obtain the posterior input estimates $\hat{\mathbf{p}}_{k|k}$ and $\mathbf{P}_{k|k}^{\mathrm{p}}$\;\vspace{1mm}
		
		Propagate the sigma points $\mathbfcal{X}_{k|k-1}^{j}$ through \cref{eq:obs-2} to obtain the predicted output $\hat{\mathbf{y}}_{k|k-1}$\;\vspace{1mm}
		
		Calculate the predicted output covariance $\mathbf{P}_k^{\,\mathrm{yy}}$ and Kalman gain using \cref{eq:obs3,eq:obs4,eq:obs5}\;\vspace{1mm}
		
		Obtain the posterior state estimates $\hat{\mathbf{x}}_{k|k}$ and $\mathbf{P}_{k|k}$ using \cref{eq:UKF-update-1a,eq:UKF-update-1b}\;\vspace{1mm}
	}
\end{algorithm}

One possibility of integrating the input estimation problem in the Bayesian framework is to simply augment $\tilde{\mathbf{x}}_k$ with $\mathbf{p}_k$, which would result to twice-augmented state, containing both parameters and inputs. An alternative to this, which was initially proposed for input-state estimation in linear systems \cite{EftekharAzam2015} and subsequently extended to input-state-parameter problems in \cite{Dertimanis2019}, is the combination of \cref{eq:input-process} with the observation equation, namely \cref{eq:augmented-output}, in order to form a second state-space model, in which $\mathbf{p}_k$ is considered the state and $\mathbf{x}_k$ is treated as a constant. This results in a dual filtering problem, which is recursively solved, first for the input, using a standard Kalman filter, and subsequently for the augmented state, 
\color{edits} using a particle-based filter, at each time step. In this sense, the augmented problem, which refers to the state and parameter estimation, can be practically solved by any of the algorithms documented in \cref{sec:state-parameter} however, for the sake of simplicity and computational efficiency, the dual filter will be presented in this paper in terms of the UKF. This results to the so called Dual Kalman Filter - Unscented Kalman Filter (DKF -UKF) \cite{Dertimanis2019}, whose implementation is in detail documented in Algorithm \ref{alg:DKF-UKF}.

It should be underlined that the problem of concurrent input, state and parameter estimation has been only recently addressed by the research community using recursive Bayesian techniques however, a number of alternatives to the DKF-UKF is already available. A straightforward approach for dealing with such a problem consists in augmenting the state with both input and parameter vectors \cite{Naets2015} and subsequently performing the sequential inference by means of a linearized or particle-based filter. When a short time delay is allowed in the estimated quantities, the inference may be carried out using the smoothing algorithm presented in \cite{Maes2019a}, which is essentially an extension of the input-state estimator proposed in \cite{Gillijns2007a}. Lastly, the problem can be solved again in a fully Bayesian context using a Gaussian Process Latent Force Model (GPLFM) for the modeling of the forcing term dynamics, in conjunction with Markov Chain Monte Carlo (MCMC) for the inference of unknown system parameters \cite{Rogers2020a}.

The main difference among these alternatives is associated with the input estimation step, which is based on different assumptions and formulations for each filter. In the dual schemes, such as DKF-UKF, the transition model of the input dynamics is used to form a second state space, which results in two filtering steps; one for the state and one for the input. The main feature of such a formulation is the decoupling of the error covariance matrices related to the input and state processes, which in turn implies that input corrections are not dependent on the state updates. On the contrary, a state augmentation approach would integrate the input process to a single augmented state-space model, thus resulting to a coupled error covariance matrix and interdependence between input and state corrections. In both of these formulations, the input noise covariance matrices, which essentially regulate the range of allowable input fluctuations, need be specified. This step is substituted by the kernel related parameters when using the Gaussian Process Latent Force Model for the input dynamics, while it is completely bypassed when using the smoothing algorithm proposed in \cite{Maes2019a}, which is characterized by a more robust performance.

\color{black}

\section{Applications}
\label{sec:applications}

\begin{figure}[t!]
	\centering
	\includegraphics[width=0.475\textwidth]{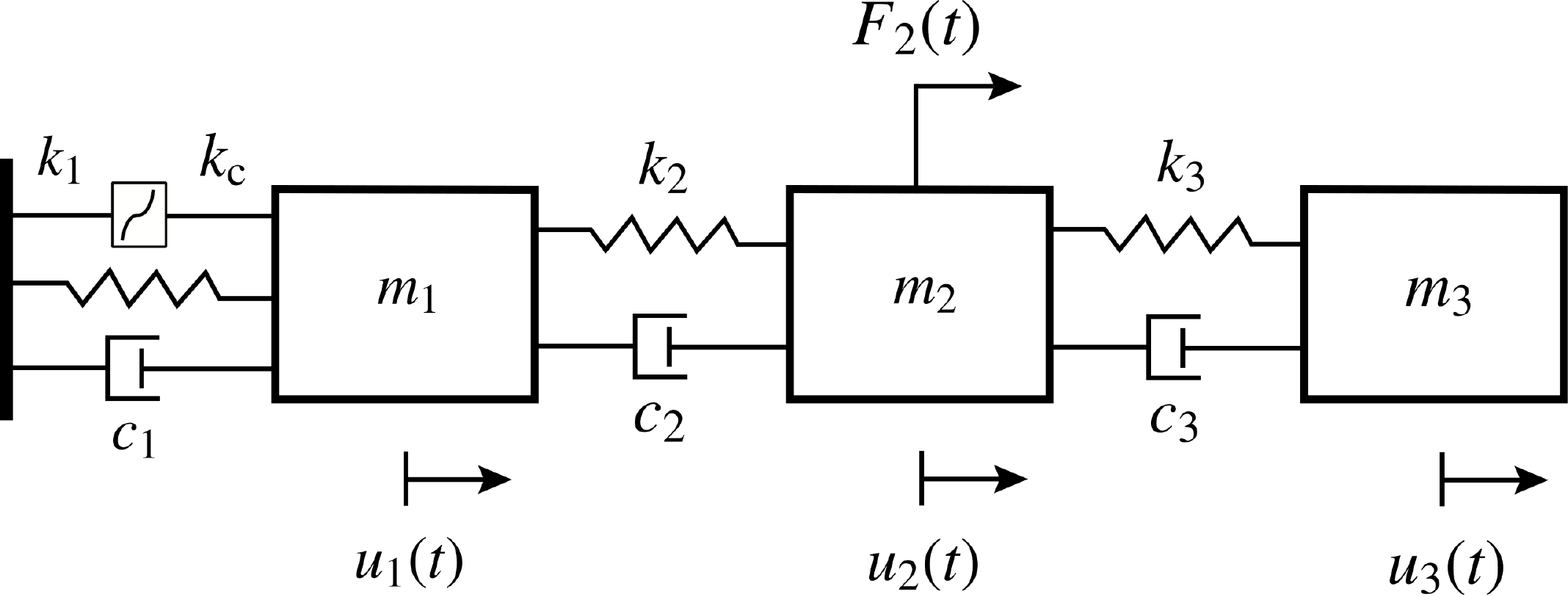}
	\caption{Model of the 3-degree-of-freedom spring-mass-damper system in which the fist mass is connected with a nonlinear spring}
	\label{fig:3dof-system}
\end{figure}

In this section, the presented algorithms for recursive Bayesian estimation are tested and compared using an academic example which consists of a 3-degree-of-freedom spring-mass-damper system, as shown in \cref{fig:3dof-system}. The three masses $m_1$, $m_2$ and $m_3$, which are all equal to $1000$ kg, are connected to dampers $c_1$, $c_2$ and $c_3$, equal to $250$ Ns/m, and linear springs $k_1$, $k_2$ and $k_3$, whose stiffness is equal to $10.000$ N/m. The first mass $m_1$ is further connected to a nonlinear spring with $k_c=10^{12}$ N/m, whose restoring force is a cubic function of the corresponding displacement $u_1(t)$. The system is lastly subjected to a white noise excitation force which is applied to $m_2$, as shown in \cref{fig:3dof-system}.

\color{edits}
The dynamic motion of the system in the continuous-time domain is described by the following nonlinear second order differential equation
\begin{equation}
	\mathbf{M}\?\ddot{\mathbf{u}}(t) + 
	\mathbf{r}\?\big(\mathbf{u}(t), \dot{\mathbf{u}}(t)\big) = \mathbf{p}(t)
	\label{eq:eom}
\end{equation}
where $\mathbf{u}(t) = \big[u_1(t)\ \, u_2(t)\ \, u_3(t)\big]^{\mathrm{T}}$ is the displacement vector, $\mathbf{M}\in\mathbb{R}^{3\times 3}$ is the mass matrix, $\mathbf{r}\?\big(\mathbf{u}(t), \dot{\mathbf{u}}(t)\big)\in\mathbb{R}^{3}$ is the vector of restoring forces that depends on the velocity and displacement terms and $\mathbf{p}(t)\in\mathbb{R}^{3}$ is the vector of external forces. Upon substitution of the system matrix expressions, \cref{eq:eom} takes the following form

\begin{equation*}
	\underbracket{
	\begin{bmatrix}
		m_1 & 0 & 0\\
		0 & m_2 & 0\\
		0 & 0 & m_3
	\end{bmatrix}}_{\mathbf{M}}
	\underbracket{
	\begin{bmatrix}
		\ddot{u_1}\\
		\ddot{u_2}\\
		\ddot{u_3}
	\end{bmatrix}}_{\mathbf{\ddot{u}}(t)}
	+
	\underbracket{
	\begin{bmatrix}
		c_1+c_2 & -c_2 & 0\\
		-c_2 & c_2+c_3 & -c_3\\
		0 & -c_3 & c_3
	\end{bmatrix}}_{\mathbf{C}}
	\underbracket{
	\begin{bmatrix}
		\dot{u_1}\\
		\dot{u_2}\\
		\dot{u_3}
	\end{bmatrix}}_{\dot{\mathbf{u}}(t)}
	+
	\underbracket{
	\begin{bmatrix}
		k_1+k_2 & -k_2 & 0\\
		-k_2 & k_2+k_3 & -k_3\\
		0 & -k_3 & k_3
	\end{bmatrix}}_{\mathbf{K}}
	\underbracket{
	\begin{bmatrix}
		u_1\\
		u_2\\
		u_3
	\end{bmatrix}}_{\mathbf{u}(t)}
	+
	\underbracket{
	\begin{bmatrix}
		k_c\cdot u_1^3\\
		0\\
		0
	\end{bmatrix}}_{\mathbf{r}_{\mathrm{nl}}\left(\mathbf{u}(t) \right)}
	=
	\underbracket{
	\begin{bmatrix}
		0\\
		F_2\\
		0
	\end{bmatrix}}_{\mathbf{p}(t)}
	\label{eq:3dof-eom}
\end{equation*}
in which the explicit dependency on time is omitted for the sake of simplicity and the restoring force term is expanded and written as the summation of the damping-related, linear stiffness and nonlinear stiffness contributions. 

\color{black}

\begin{figure}[b!]
	\centering
	\begin{subfigure}[b]{0.49\textwidth}
		\includegraphics[width=\textwidth]{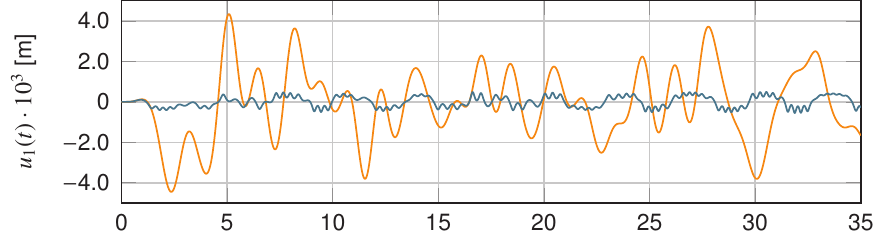}
		\includegraphics[width=\textwidth]{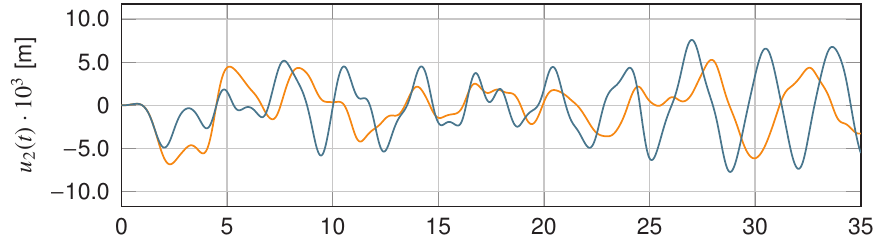}
		\includegraphics[width=\textwidth]{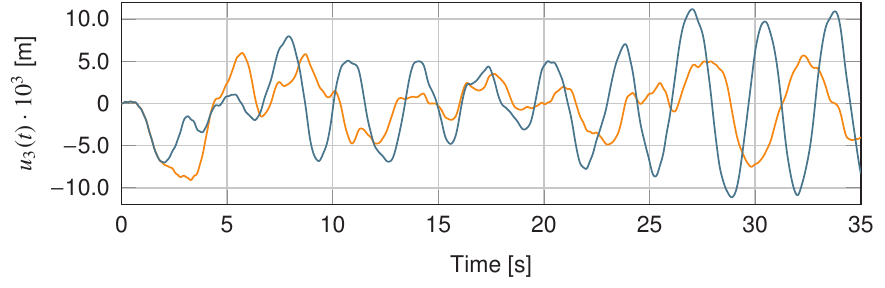}
	\end{subfigure}
	\begin{subfigure}[b]{0.49\textwidth}
		\includegraphics[width=\textwidth]{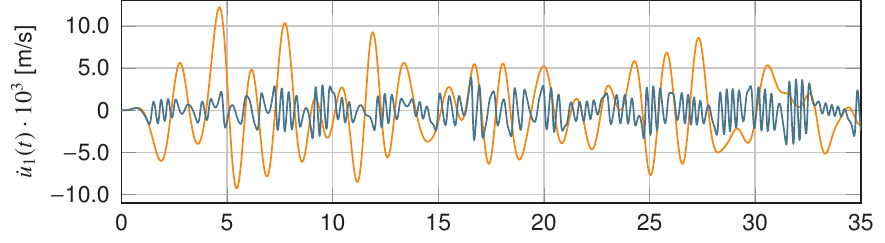}
		\includegraphics[width=\textwidth]{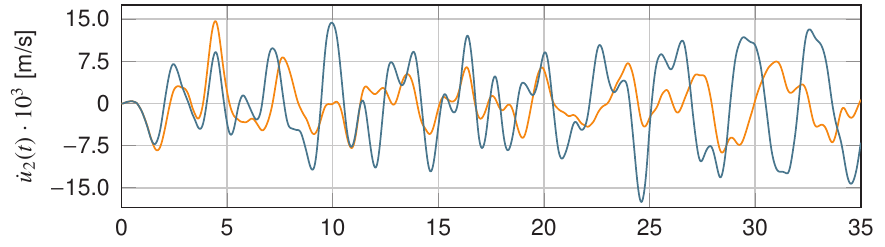}
		\includegraphics[width=\textwidth]{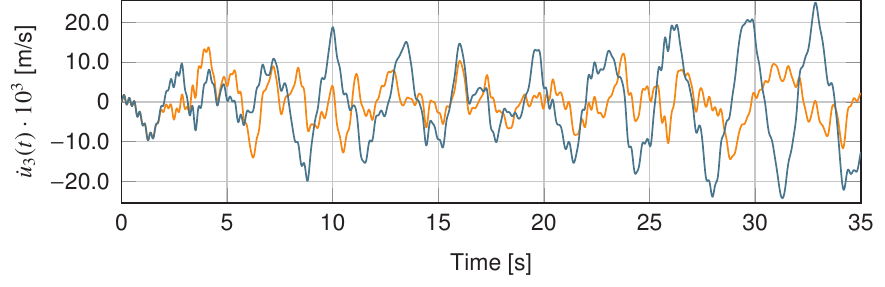}
	\end{subfigure}\vspace{-2mm}
	\caption{Comparison of the linear (orange) and nonlinear (blue) state response}
	\label{fig:linear-nonlinear}
\end{figure}

\begin{figure}[h]
	\centering
	\includegraphics[width=0.48\textwidth]{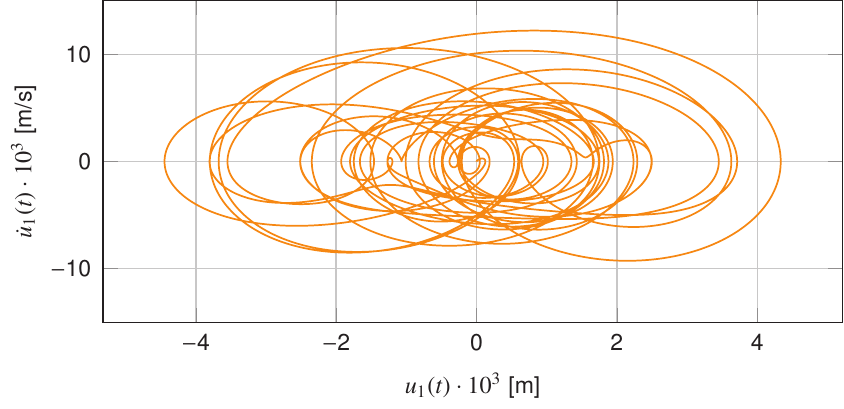}
	\hfill
	\includegraphics[width=0.48\textwidth]{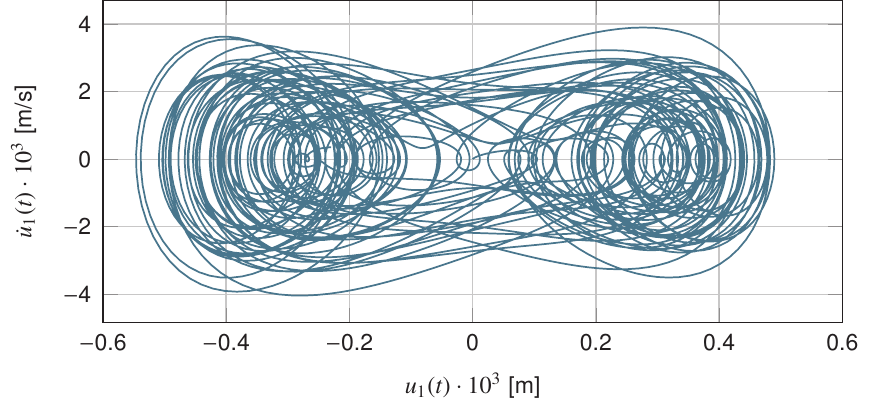}
	\caption{Phase portrait of the first degree of freedom for the linear (left) and nonlinear (right) system}
	\label{fig:phase-portrait}
\end{figure}

In order to highlight the effect of nonlinearity, the system is initially simulated as a linear one, in which all three masses are connected among them with springs $k_1$, $k_2$ and $k_3$, and subsequently as a nonlinear one, in which the cubic spring $k_c$ is further included, using the same excitation. The difference in the simulated states is plotted in \cref{fig:linear-nonlinear}, where the vibration amplitude of the first mass is significantly reduced, in terms of both velocity and displacement, and the majority of energy is mainly transferred to the third mass, whose amplitude is seen to increase. The difference in the dynamics of the two systems is also evidenced by the phase space trajectories of the first mass which are presented in \cref{fig:phase-portrait}.

\color{edits}

The problem of Bayesian inference in structural dynamics is presented in the following sections in the form of state, state-parameter and state-input-parameter estimation, with the aiming of calculating the posterior expectation described by \cref{eq:condmean}, when $\mathbf{g}(\mathbf{x}_k) = \mathbf{x}_k$. These quantities are oftentimes of interest, not only in structural dynamics, but also in SHM applications. Apart from the evident case of retrieving the dynamic response at unmeasured  locations, the inference of these quantities may serve for identifying unknown system parameters \cite{Erazo2018,Calabrese2018}, as well as material constitutive models \cite{Ebrahimian2015}. In the context of real-time vibration mitigation for structural systems, the filtering algorithms can be further combined with control schemes \cite{Miah2015}, in which case the aim is to deliver the inferred quantities of interest to the controller. Lastly, the problem of damage detection in structural and mechanical systems can be also tailored to a sequential Bayesian inference problem. Not only in terms of damage in the form of cracks \cite{Tatsis2022}, but also in terms of sensor-faults \cite{Kobayashi2003} and accumulated fatigue damage \cite{Cadini2017}.

\begin{figure}[h]
	\centering
	\includegraphics[width=0.9\textwidth]{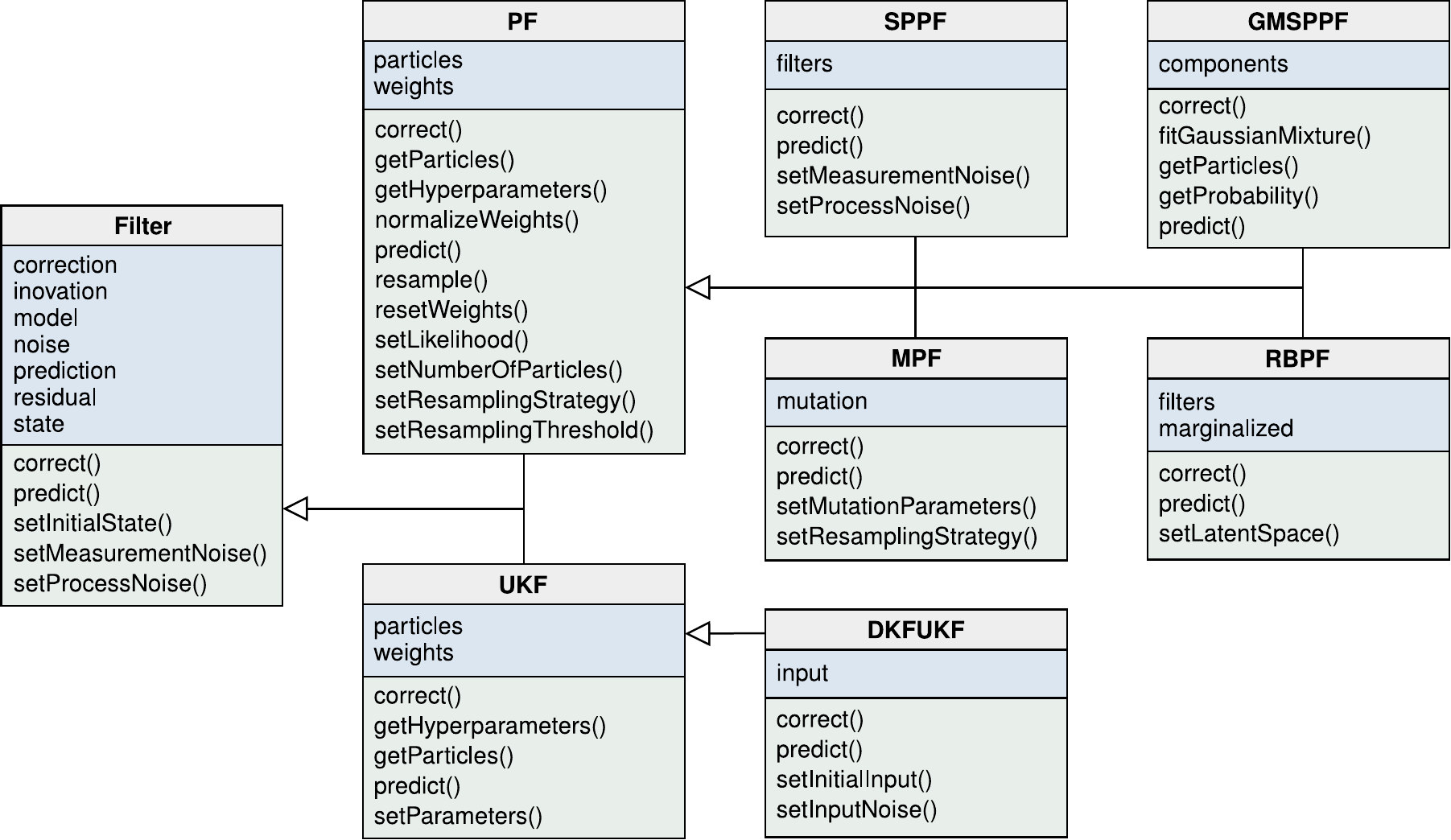}
	\caption{Class diagram of the python library for Sequential Bayesian Inference}
	\label{fig:uml}
\end{figure}

The steps of Sequential Bayesian Inference are performed using the associated open access python library, which is available via \href{https://github.com/ETH-WindMil/JSD-SBI}{GitLab}. The structure of the object-oriented library is displayed in the UML (Unified Modeling Language) class diagram shown in \cref{fig:uml}, while each class is individually documented in the repository. The algorithms presented in this paper inherit the attributes and methods of the abstract parent class named \texttt{Filter}, which encapsulates all common attributes and methods, contained in the blue and green boxes of the class diagram, respectively. Apart from the UKF and DKF-UKF algorithms, which are based on the Unscented Transform and therefore make use of deterministically placed samples, all algorithms inherit their attributes and methods from the PF, which is illustrated in \cref{fig:uml} by the extension arrows of the class diagram. The initialization of the algorithms is performed by the \texttt{setInitialState()} method, while the process and measurement noise covariances are specified by the \texttt{setProcessNoise} and \texttt{setMeasurementNoise} methods. The main inference steps of prediction and correction are carried out using the methods \texttt{correct()} and \texttt{predict()}, which are implemented by all filters accordingly.

\color{black}

The results reported in the following sections are obtained by means of simulated data, which are generated according to the above-described nonlinear set-up. In all cases, the vibration response of the system is measured in terms of displacements in $m_1$ and accelerations in $m_1$, $m_2$ and $m_3$. The simulated response in terms of these quantities is obtained by means of a 4th order Runge-Kutta integration scheme, subsequently polluted with a 3\% which Gaussian noise, and lastly combined with the filtering algorithms for the solution of state, state-parameter and input-state-parameter estimation problems. It should be underlined, that the results presented below are obtained upon proper tuning of the noise covariance matrices, which constitutes a decisive step for optimal estimation performance.

\color{edits}

In this paper, the noise terms are manually tuned by minimizing the difference between measured signals, which are not used in the filtering equations, and the corresponding posterior signals. This is a standard practice in the literature for selecting the covariance values \cite{Maes2019a,Dertimanis2019}, which are used to account for the uncertainties related to the state, measurement, input and parameter equations. These uncertainties are propagated through the filtering equations and finally reflected on the estimated probability densities. This implies that an increased uncertainty in the process noise $\mathbf{v}_k$, which can be specified by the corresponding covariance matrix $\mathbf{Q}_k$, would allow strong fluctuations of the state vector $\mathbf{x}_k$ around the model predicted values, while a reduced uncertainty would force the state dynamics to be strictly governed by the underlying equations. Under this perspective, the covariance matrix of the input process is typically tuned to high values, so as to allow for strong variations of the external loads, such as sudden impacts. A formal way of tuning this noise term is the L-curve \cite{Hansen1992}, which is extracted though from an offline step. On the other hand, a low value is assigned to the covariance matrix of the unknown parameters, thus allowing only slow variations.

Apart from the agreement between measurements and predictions, the tuning point can be selected on the basis of different optimality metrics \cite{Matisko2012}, such as the whiteness of the residual sequence. The numerous contributions in this regard can be divided into two categories, the ones operating online, thus delivering an adaptive noise estimation, and the ones extracting the estimates on the basis of data batches. A straightforward approach falling into the first category, which can be also tailored to the algorithms presented in this paper, consists in augmenting the state vector with the unknown noise parameters and solving a joint state-parameter problem \cite{Kontoroupi2016}. An alternative online estimate can be obtained using a covariance matching approach \cite{Gao2015}, which aims at estimating the noise covariance matrices using the residuals of the inference step. When relaxing the constraint of online performance, noise estimates can be obtained by means of the Autocovariance Least-Squares technique \cite{Odelson2006,Rajamani2009} or using an Expectation Maximization (EM) approach \cite{Ge2017}. It should be underlined that the topic of optimal noise identification is out of the scope of this paper however, the reader is referred to \cite{Dunik2017} for a more thorough review and comparison of noise estimation methods.

\color{black}


\subsection{State estimation}
\label{sub:state}

\color{black}

In this first case study,
\color{edits}
which deals with the simplest form of inference in dynamic systems
\color{black}
, the unobserved state of the structure shown in \cref{fig:3dof-system} is estimated using the UKF, PF and SPPF.
\color{edits}
To do so, the system dynamics described by \cref{eq:3dof-eom} must be written in a state-space form in accordance with \cref{eq:state,eq:obs}. This requires the construction of the state and measurement functions $\mathbf{F}$ and $\mathbf{H}$, which are obtained upon definition of the state vector $\mathbf{x}(t) = \vect\big(\left[\mathbf{u}(t)\ \, \dot{\mathbf{u}}(t)\right]\big)$, as follows
\begin{align*}
	\mathbf{F}\big(\mathbf{x}(t), \mathbf{p}(t)\big) & = 
	\begin{bmatrix}
		\dot{\mathbf{u}}(t)\\[1mm]
		-\mathbf{M}^{-1}\Big(\mathbf{p}(t)-\mathbf{C}\dot{\mathbf{u}}(t) - \mathbf{K}\mathbf{u}(t) -\mathbf{r_{\mathrm{nl}}\big(\mathbf{u}(t) \big)}\Big)\\[1mm]
	\end{bmatrix}\\[5mm]
	\mathbf{H}\big(\mathbf{x}_k, \mathbf{p}_k \big) & = 
	\begin{bmatrix}
		\mathbf{S}_{\mathrm{d}}\,\mathbf{u}_{k}\\[1mm]
		-\mathbf{M}^{-1}\Big(\mathbf{p}_k-\mathbf{C}\dot{\mathbf{u}}_k - \mathbf{K}\mathbf{u}_k -\mathbf{r_{\mathrm{nl}}}\big(\mathbf{u}_k\big)\Big)\\[1mm]
	\end{bmatrix}
\end{align*}

\begin{figure}[t!]
	\centering
	\begin{subfigure}[b]{0.49\textwidth}
		\includegraphics[scale=0.95]{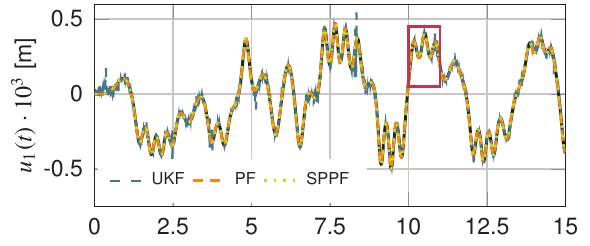}
		\hspace{-5mm}
		\includegraphics[scale=0.95]{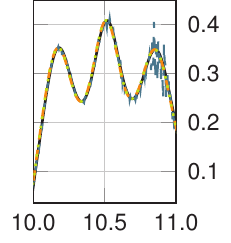}
		\includegraphics[scale=0.95]{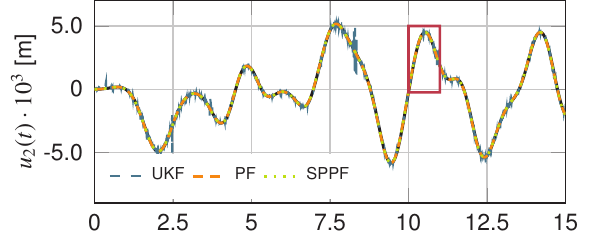}
		\hspace{-5mm}
		\includegraphics[scale=0.95]{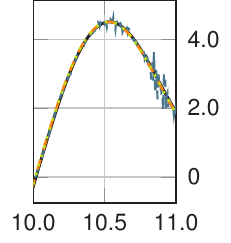}\vspace{-1.5mm}
		\includegraphics[scale=0.95]{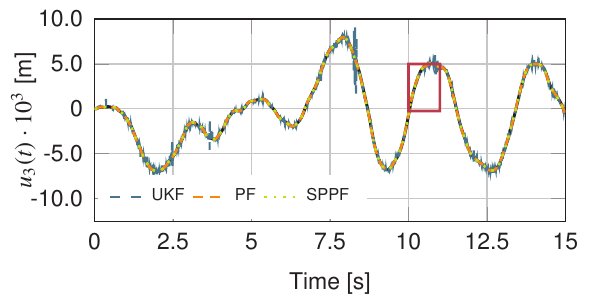}
		\hspace{0mm}
		\includegraphics[scale=0.95]{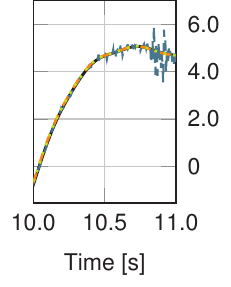}
	\end{subfigure}\hfill
	\begin{subfigure}[b]{0.49\textwidth}
		\includegraphics[scale=0.95]{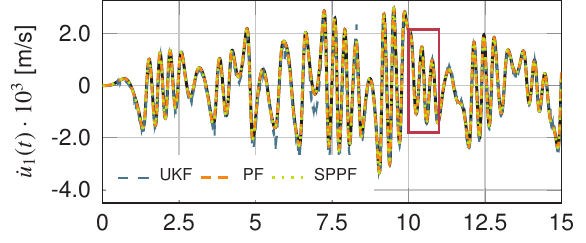}
		\hspace{-4mm}
		\includegraphics[scale=0.95]{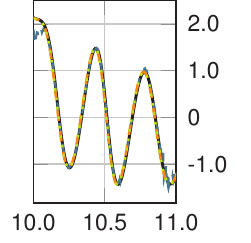}
		\includegraphics[scale=0.95]{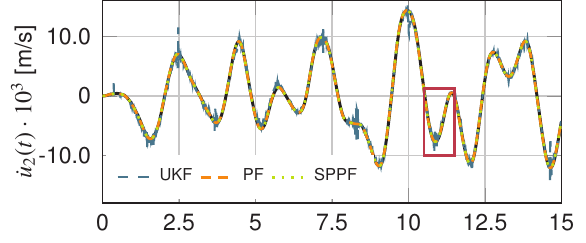}
		\hspace{-4mm}
		\includegraphics[scale=0.95]{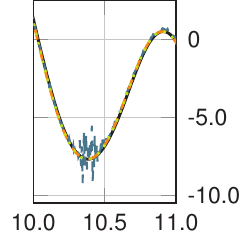}
		\includegraphics[scale=0.95]{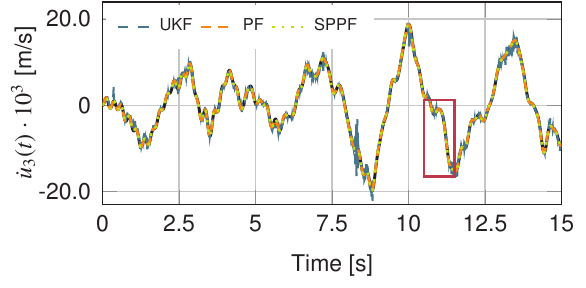}
		\hspace{-1.5mm}
		\includegraphics[scale=0.95]{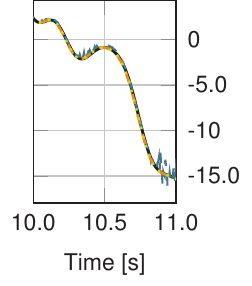}
	\end{subfigure}
	\caption{Comparison of the UKF (blue), PF (orange) and SPPF (green) state estimates against the ground truth (black)}
	\label{fig:state-estimation}
\end{figure}

\noindent
where $\mathbf{S}_{\mathrm{d}}$ is a Boolean matrix selecting the observed displacements, which for this case study reads $\mathbf{S}_{\mathrm{d}} = \left[0\ \, 1\ \, 0 \right]$. It should be noted that in order to perform the inference steps of each algorithm in the discrete-time domain, the state function should be further discretized, according to \cref{eq:integ}. This step is herein carried out using a 4th order Runge-Kutta scheme and yields function $\mathbf{G}(\mathbf{x}_k, \mathbf{p}_k)$, which is the discrete-time equivalent of $\mathbf{F}\left(\mathbf{x}(t), \mathbf{p}(t)\right)$.

\color{black}

The UKF is based on a deterministic sampling of the sought-after distributions, using $2n+1$ sigma points, where $n$ designates the size of the state vector. For the considered example, this results to a total of 13 particles, which are symmetrically placed with respect to the prior state statistics. The PF is herein based on $N=30$ particles, which is the required number of samples for achieving a sufficiently good performance, comparable to the one delivered by the UKF, while the estimates of the SPPF are based on 25 particles. The estimates obtained from the three algorithms are plotted in \cref{fig:state-estimation} in terms of displacements and velocities at all three degrees of freedom of the system, with blue, orange and green lines. The target response is represented in all figures by a continuous black line, which is only partially visible due to the good agreement between the actual and predicted time histories.

In all three runs, namely UKF, PF and SPPF, the state is initialized with a zero mean $\hat{\mathbf{x}}_{0|0}=\mathbf{0}_n$ and covariance matrix $\mathbf{P}_{0|0} = 10^{-20}\cdot\mathbf{I}_n$, where $\mathbf{0}_n$ and $\mathbf{I}_n$ denote the null and identity matrices of size $n$, respectively. Similarly, identical measurement and process noise covariance matrices are used for all three algorithms, equal to $\mathbf{R}_k = \diag\left(\left[10^{-5}, \ 10^{-8}, \ 10^{-8}, \ 10^{-8}\right]\right)$ and $\mathbf{Q}_k = \diag\left(\left[10^{-9}, \ 10^{-9}, \ 10^{-9}, \ 10^{-14}, \ 10^{-14}, \ 10^{-14}\right]\right)$ respectively. The UKF distribution-related parameters are set to the default values, namely $\alpha=1$ and $\beta=2$, while the effective sample size of the PF and SPPF, which essentially represents the resampling limit, is $N_{\mathrm{eff}}=0.2\cdot N$, with $N$ denoting the number of particles.


\subsection{State and parameter estimation}
\label{sub:state-parameter}

\color{edits}

In the previous section, it was assumed that the parameters of the state-space model were known, which is not always the case in practical applications. As underlined in the beginning of \cref{sec:applications}, a typical application of sequential Bayesian inference in structural dynamics systems pertains to the identification of uncertain or unknown system parameters. Within this context, it is further assumed in this section that the nonlinear spring stiffness is an unknown system parameter and will be estimated alongside the unobserved state, using measurements of the vibration response. It should be clarified that it is only the value of $k_c$ that is to be estimated  and not the entire model class, namely the form of nonlinearity. 

To do so, the state augmentation technique is used, whereby the system dynamics are written in an augmented state-space form, which is obtained by appending the unknown parameters to the state vector. In this case study, the parameter to be estimated is the value of the nonlinear stiffness term, so that $\bm{\theta} = k_c$ and $\tilde{\mathbf{x}}(t) = \vect\big(\left[\mathbf{u}(t),\, \dot{\mathbf{u}}(t),\, \bm{\theta}\right]\big)$. As pointed out in \cref{sec:state-parameter}, the transition model of the parameter vector is assumed to be a Gaussian random walk, which in the continuous-time domain is translated into the following equation: $\dot{\bm{\theta}} = \mathbf{0}$. This model enables the construction of the continuous-time state function, which reads
\begin{equation*}
	\mathbf{F}\big(\tilde{\mathbf{x}}(t), \mathbf{p}(t)\big) = 
	\begin{bmatrix}
		\dot{\mathbf{u}}(t)\\[1mm]
		-\mathbf{M}^{-1}\Big(\mathbf{p}(t)-\mathbf{C}\dot{\mathbf{u}}(t) - \mathbf{K}\mathbf{u}(t) -\mathbf{r_{\mathrm{nl}}\big(\mathbf{u}(t) \big)}\Big)\\[1mm]
		0\\
	\end{bmatrix}
\end{equation*}
and is transformed to the discrete-time equivalent $\tilde{\mathbf{G}}\left(\tilde{\mathbf{x}}_k, \mathbf{p}_k \right)$ by means of a 4th order Runge-Kutta scheme. The observation function $\mathbf{H}\left(\mathbf{x}_k, \mathbf{p}_k \right)$ is identical to the one used in the previous case study since the same response quantities are observed.

\color{black}

\begin{figure}[t!]
	\centering
	\begin{subfigure}[b]{0.49\textwidth}
		\includegraphics[scale=0.95]{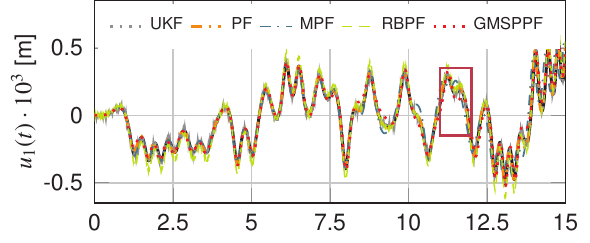}
		\hspace{-5mm}
		\includegraphics[scale=0.95]{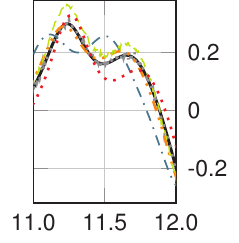}\vspace{-1.5mm}
		\includegraphics[scale=0.95]{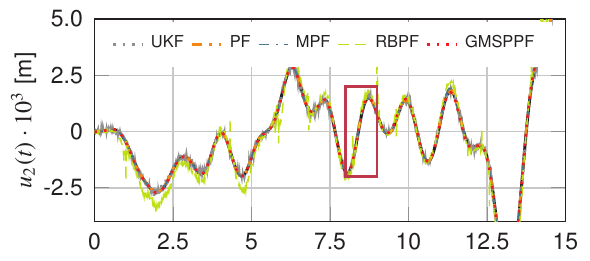}
		\hspace{-5mm}
		\includegraphics[scale=0.95]{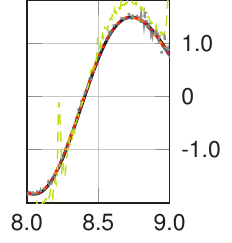}
		\includegraphics[scale=0.95]{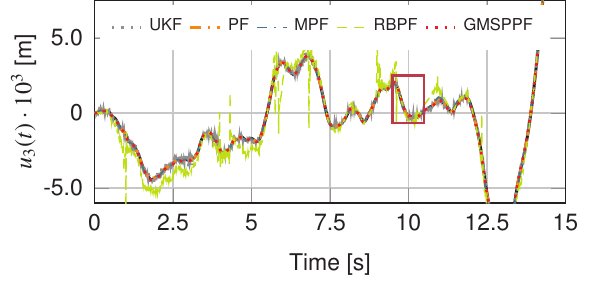}
		\hspace{0mm}
		\includegraphics[scale=0.95]{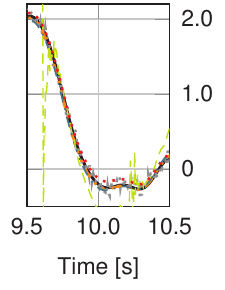}
	\end{subfigure}\hfill
	\begin{subfigure}[b]{0.49\textwidth}
		\includegraphics[scale=0.95]{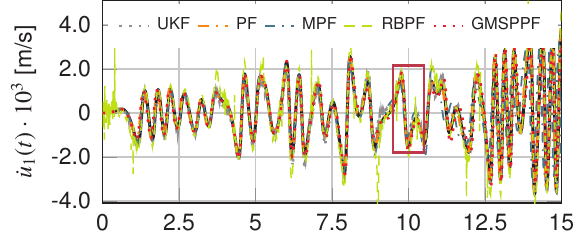}
		\hspace{-5mm}
		\includegraphics[scale=0.95]{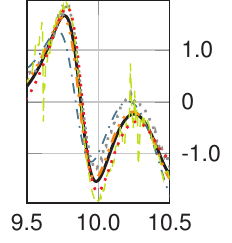}
		\includegraphics[scale=0.95]{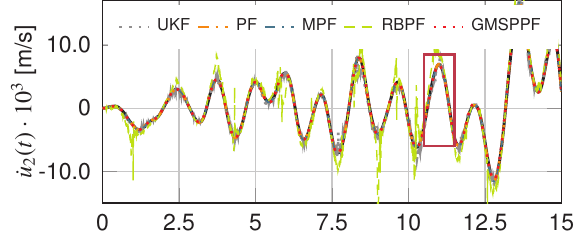}
		\hspace{-5mm}
		\includegraphics[scale=0.95]{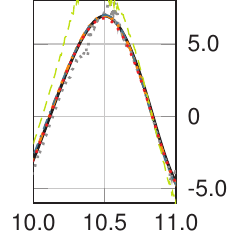}
		\includegraphics[scale=0.95]{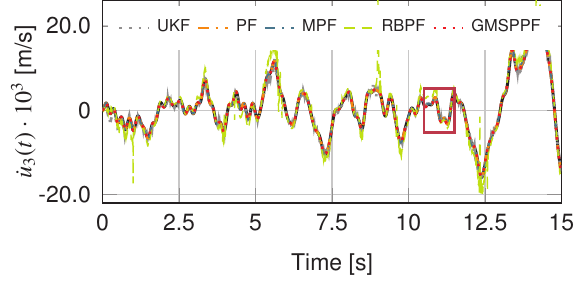}
		\hspace{-0.5mm}
		\includegraphics[scale=0.95]{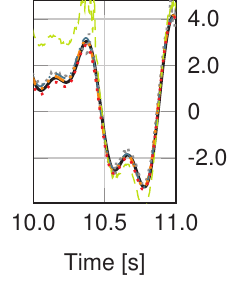}
	\end{subfigure}
	\caption{Comparison of the UKF (gray), PF (orange), MPF (blue), RBPF (green) and GMSPPF (red) state estimates against the ground truth (black)}
	\label{fig:state-parameter-1}
\end{figure}

In this case study, the use of UKF results to the most computationally efficient solution of the Bayesian problem using only 15 sigma points, while the remaining algorithms, namely PF, MPF, GMSPPF and RBPF are based on a significantly larger amount of particles.
\color{edits}
In the latter algorithm, the states related to the dynamics of the system are marginalized, so that $\mathbf{x}_k^{\mathrm{b}} = \vect\left(\left[\mathbf{u}_k,\  \dot{\mathbf{u}}_k \right] \right)$, and their posterior is obtained by means of a UKF, while the PF sampling takes place only for the sought-after parameter, $\mathbf{x}_k^{\mathrm{a}} = k_c$.
It should be noted that major advantage of the RBPF is obtained when the filtering equations of the marginalized state can be analytically computed. When this condition is not met, which might be due to nonlinearities existing in the state and/or observation equations, the inference over the marginalized state can be performed by replacing the analytical evaluation with any filter capable of accounting for nonlinearities, such as the EKF or the UKF \cite{Sarkka2007}. Consequently, the use of UKF for $\mathbf{x}_k^{\mathrm{b}}$ is herein owed to the fact that both state and observation equations are characterized by nonlinear terms.

\color{black}

\begin{figure}[h]
	\centering
	\includegraphics[scale=1]{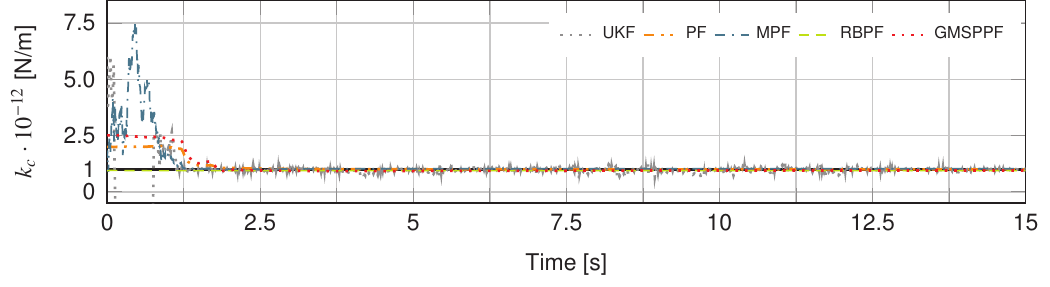}
	\caption{Comparison of the PF (orange), MPF (blue) and RBPF (green) parameter estimates with the ground truth (black)}
	\label{fig:state-parameter-2}
\end{figure}

The state estimates of each algorithm are compared against the ground truth in \cref{fig:state-parameter-1}, while the corresponding parameter estimates are displayed in \cref{fig:state-parameter-2}. These are obtained upon initialization of all three filters with a zero state $\hat{\mathbf{x}}_{0|0}$ and a diagonal covariance matrix $\mathbf{P}_{0|0}$, whose entries are equal to $10^{-11}$, except for the one corresponding to the parameter $k_c$ which is set to $10^{-2}$. The measurement noise covariance for the UKF is $\mathbf{R} = \diag\left(\left[10^{-9},\ 10^{2},\ 10^{-3},\ 10^{-3}  \right] \right)$, for PF: $\mathbf{R} = \diag\left(\left[10^{-9},\ 10^{1},\ 10^{-3},\ 10^{-3}  \right] \right)$, for the MPF: $\mathbf{R} = \diag\left(\left[10^{-9},\ 10^{2},\ 10^{-2},\ 10^{-2}  \right] \right)$, for the RBPF: $\mathbf{R} = \diag\left(\left[10^{-14},\ 10^{-4},\ 10^{-6},\ 10^{-4} \right] \right)$ and for the GMSPPF: $\mathbf{R} = \diag\left(\left[10^{-9},\ 10^{1},\ 10^{-3},\ 10^{-2} \right] \right)$, while an identical process noise covariance is used for all five algorithms: $\mathbf{Q}=\diag\left(\left[10^{-11},\ 10^{-9},\ 10^{-9},\ 10^{-11},\ 10^{-9},\ 10^{-9},\ 10^{-25} \right] \right)$.
\color{edits}
The last entry of the process noise is assigned to the unknown parameter, which as discussed in \cref{sec:applications} is chosen such that only slow variations of the sought-after parameters are allowed.
\color{black}

In what concerns the filtering parameters, the PF and GMSPPF in this example run with $N=1000$ particles, whose resampling threshold is set to $N_{\mathrm{eff}}=0.2\cdot N$, while the MPF is seen to achieve a similar accuracy in terms of both state and parameter estimates using $N=400$ particles with $N_{\mathrm{eff}}=0.3\cdot N$, $p_{\mathrm{r}}=0.05$ and $p_{\mathrm{m}}=0.25$. This difference in the required number of particles is owed to the mutation scheme, which introduces a more efficient exploration of the parameter space and creates a significant computational advantage for the MPF when it comes to constant parameter identification, as has been also observed in \cite{Chatzi2013}. This effect is also evidenced in \cref{fig:state-parameter-2}, which depicts the estimated parameter time histories. Although all three algorithms converge to the target parameter value, the MPF is seen to have a faster convergence with respect to the PF, which is also combined with a more extensive exploration of the parameter space. On the other hand, the RBPF achieves a quite fast convergence to the target parameter value, using only 70 particles for the non-marginalized part of the state vector and a resampling threshold $N_{\mathrm{eff}}=0.2\cdot N$.

\subsection{Input, state and parameter estimation}
\label{sub:input-state-parameter}

A further relaxation to the optimal Bayesian estimation problem is introduced by assuming that the system dynamics are driven by an unmeasured and unknown input excitation, which can be also estimated along with the state and system parameters. The DKF-UKF scheme, which was initially proposed in \cite{Dertimanis2019} for linear systems and presented herein in Algorithm \ref{alg:DKF-UKF}, is therefore used for the input, state and parameter estimation of the nonlinear system illustrated in \cref{fig:3dof-system}. It should be noted that the filter proposed in \cite{Dertimanis2019} has been presented for linear systems, which upon augmentation become nonlinear. However, in this example, apart from the nonlinearity occurring from the state augmentation, the system contains an additional nonlinearity, which is owed to the nonlinear spring force acting on the first degree of freedom.

\begin{figure}[h]
	\centering
	\begin{subfigure}{0.49\textwidth}
		\includegraphics[scale=0.95]{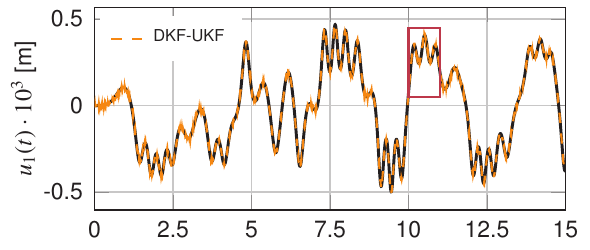}
		\hspace{-5mm}
		\includegraphics[scale=0.95]{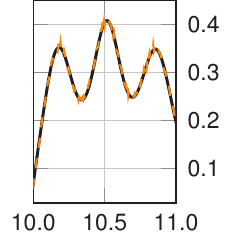}
		\includegraphics[scale=0.95]{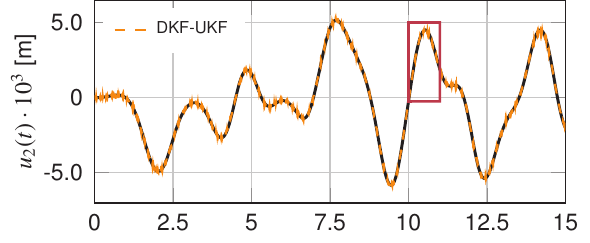}
		\hspace{-5mm}
		\includegraphics[scale=0.95]{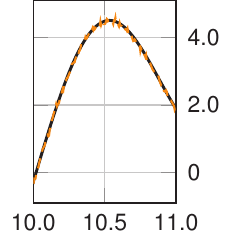}
		\includegraphics[scale=0.95]{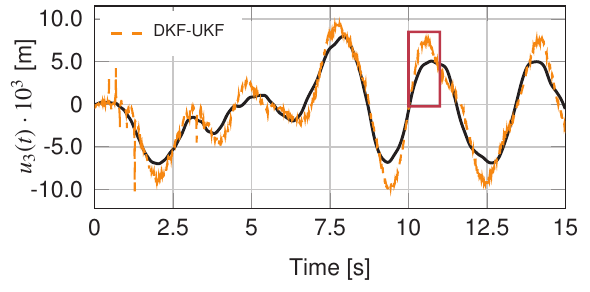}
		\hspace{0mm}
		\includegraphics[scale=0.95]{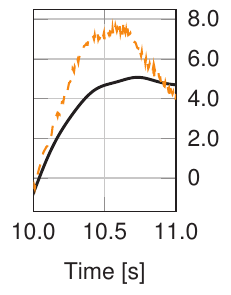}
	\end{subfigure}
	\begin{subfigure}{0.49\textwidth}
		\includegraphics[scale=0.95]{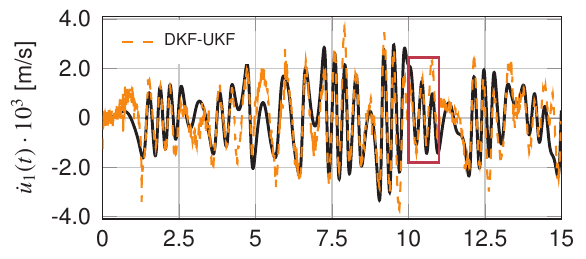}
		\hspace{-5mm}
		\includegraphics[scale=0.95]{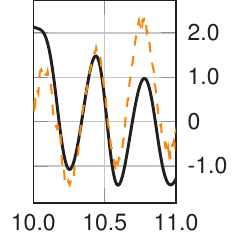}
		\includegraphics[scale=0.95]{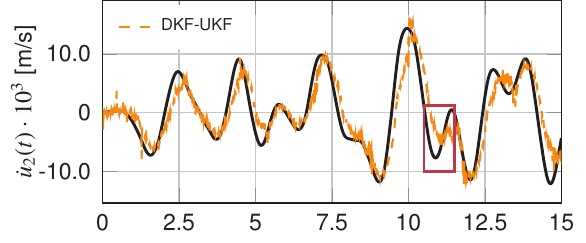}
		\hspace{-5mm}
		\includegraphics[scale=0.95]{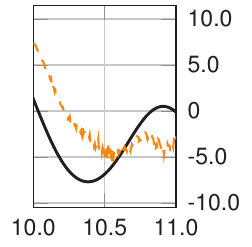}
		\includegraphics[scale=0.95]{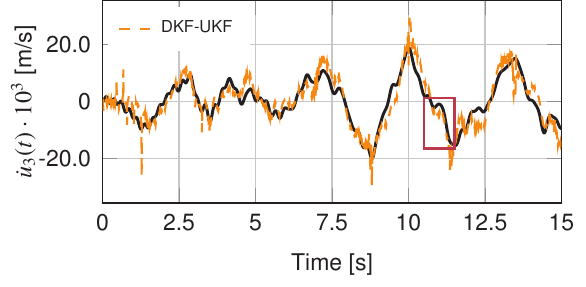}
		\hspace{-1.5mm}
		\includegraphics[scale=0.95]{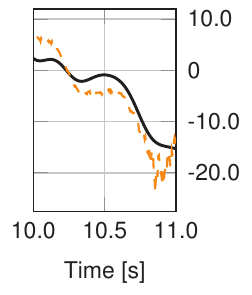}
	\end{subfigure}
	\caption{Comparison of the Dual Kalman Filter - Unscented Kalman Filter (DKF-UKF) state estimates with the ground truth (black)}
	\label{fig:input-state-parameter-1}
\end{figure}

\begin{figure}[h]
	\centering
	\includegraphics[scale=0.975]{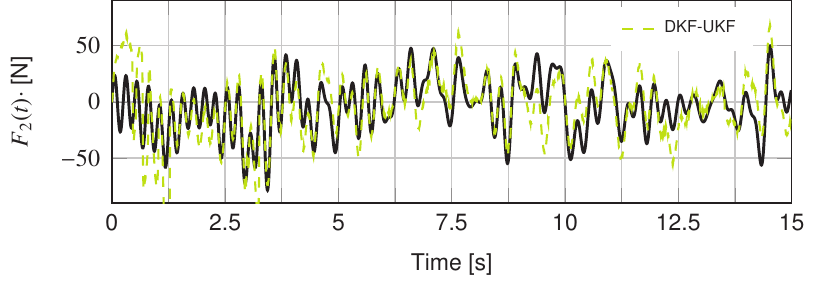}
	\hfill
	\includegraphics[scale=0.975]{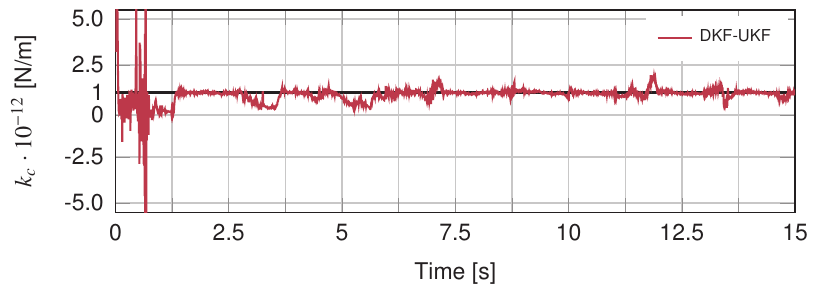}
	\caption{Comparison of the Dual Kalman Filter - Unscented Kalman Filter (DKF-UKF) input (left) and parameter (right) estimates with the ground truth (black)}
	\label{fig:input-state-parameter-2}
\end{figure}

Similarly to the state and parameter estimation problem presented in \cref{sub:state-parameter}, the dynamic equations of the system are initially augmented so as to accommodate the sought-after parameter as unknown state. Thereafter, a second stochastic process is assumed for the input evolution, as described in \cref{sec:state-input-parameter}, and the input-state-parameter estimation problem is solved using Algorithm \ref{alg:DKF-UKF}. The state estimates, along with the actual vibration response of the system are displayed in \cref{fig:input-state-parameter-1}, in which the latter is represented by a continuous black line and the DKF-UKF estimates are plotted with an orange dashed line. The corresponding input and parameter estimates, which are also delivered by the DKF-UKF are presented in \cref{fig:input-state-parameter-2}.

In terms of the algorithm parameters, a zero initial state $\hat{\mathbf{x}}_{0|0}$ is assumed, with covariance matrix $\mathbf{P}_{0|0}=\diag\left(\left[10^{-5},\ 10^{-5},\ 10^{-5},\ 10^{-5},\ 10^{-5},\ 10^{-5},\ 10^{-2} \right] \right)$ and a zero $\hat{\mathbf{p}}_{0|0}$ is also assumed for the input signal, with covariance $\mathbf{P}_{0|0}^{\,\mathrm{p}}=10^{10}\cdot\mathbf{I}_{n_{\mathrm{p}}}$. The measurement noise covariance is equal to $\mathbf{R}_k = \diag\left(\left[10^{-11},\ 10^{-9},\ 10^{-3},\ 10^{-10}  \right]\right)$ and the state process covariance matrix is tuned to the following values $\mathbf{Q}_k = \diag\left(\left[10^{-10},\ 10^{-9},\ 10^{-9},\ 10^{-8},\ 10^{-8},\ 10^{-8},\ 10^{-25} \right] \right)$. The last parameter of the DKF-UKF algorithm is the noise covariance matrix of the input process, which is equal to $\mathbf{Q}^{\mathrm{p}}_k=2\cdot 10^{\,2}\cdot \mathbf{I}_{n_{\mathrm{p}}}$.

\section{Conclusions}

This contribution provides an overview of particle-based methods for sequential Bayesian inference of nonlinear and non-Gaussian dynamic systems for Structural Health Monitoring applications. The presented class of filtering approaches is intended to tackle the difficulties associated with the nonlinearities and uncertainties encountered in typical engineering applications. As such, the formulation and specifics of a number of particle-based filtering algorithms are described along with the detailed implementation steps. The performance of the most representative recursive filters, whose implementation is available at \href{https://github.com/ETH-WindMil/JSD-SBI}{https://github.com/ETH-WindMil/JSD-SBI}, is also compared and discussed on the basis of a simple illustrative example, which is used for state, state-parameter and input-state parameter estimation.

\section*{Acknowledgements}

The authors would like to gratefully acknowledge the support of the European Research Council via the ERC Starting Grant WINDMIL (ERC-2015-StG \#679843) on the topic of "Smart Monitoring, Inspection and Life-Cycle Assessment of Wind Turbines", the ERC Proof of Concept (PoC) Grant, ERC-2018-PoC WINDMIL RT-DT on "An autonomous Real-Time Decision Tree framework for monitoring and diagnostics on wind turbines".

\bibliographystyle{unsrt}
\bibliography{library}  

\end{document}